\documentclass[12pt]{article}
\usepackage{a4}
\usepackage{amsmath}
\usepackage{amssymb}
\usepackage[dvips]{graphicx}
\def\beq{\begin{equation}}
\def\eeq{\end{equation}}
\def\beqa{\begin{eqnarray}}
\def\eeqa{\end{eqnarray}}
\def\n{\nonumber \\}

\begin{document}

\begin{flushright}
{SAGA-HE-189}\\
{KEK-TH-839}
\end{flushright}
\vskip 0.5 truecm


\begin{center} 
{\Large{\bf Chiral Anomaly  on Fuzzy 2-Sphere }}\\
\vskip 1.5cm

{\large Hajime Aoki\footnote{e-mail
 address: haoki@cc.saga-u.ac.jp}, Satoshi Iso\footnote{e-mail
 address: satoshi.iso@kek.jp} and Keiichi Nagao\footnote{e-mail
 address: nagao@post.kek.jp}}  
\vskip 0.5cm
 
$^1${\it Department of Physics, Saga University, Saga 840-8502, 
Japan  }\\

{\it and }\\

$^{2,3}${\it High Energy Accelerator Research Organization (KEK)\\
Tsukuba 305-0801, Japan}
 
\end{center}

\vskip 2cm
\begin{center} 
\begin{bf}
Abstract
\end{bf}
\end{center}

We investigate chiral anomaly for fermions in the fundamental 
representation
on noncommutative (fuzzy) 2-sphere.
In spite that this system is realized by finite dimensional 
matrices and
no regularization is necessary for either  UV or IR, 
we can reproduce the correct chiral anomaly  which is consistent
with the calculations done in flat noncommutative space.
Like the flat case, there are ambiguities to define chiral currents.
We define chiral currents
in a gauge-invariant way and a gauge-covariant way, 
and show that the corresponding
anomalous chiral Ward--Takahashi identities 
take different forms.
The Ward--Takahashi identity for the  
gauge-invariant current contains
explicit nonlocality while that for the covariant one is
given by a local expression.

\newpage
\setcounter{footnote}{0}
\section{Introduction}
\setcounter{equation}{0}
Noncommutative field theory has attracted much interest recently 
since it was realized that noncommutative geometry appears
naturally from string theory in $B_{\mu \nu}$ background\cite{SW}.
Furthermore noncommutative geometry can appear as natural background
space-time in matrix models\cite{CDS,NCMM,MLi}
that have been proposed as 
a nonperturbative formulaion of superstring theory.
Various novel properties\cite{NC}, such as gravity-like 
induced interactions
between objects in noncommutative space,
open Wilson lines, UV/IR duality, 
bi-locality and background independence, demonstrate that  
Yang-Mills theory in noncommutative space-time
may be more appropriately interpreted as a stringy theory
than an ordinary local field theory.
Since noncommutative field theories can be formulated 
as matrix models, these novel properties support fundamentality
of matrix models as a constructive formulation of 
superstring\cite{IKKT}.

Anomalies of noncommutative gauge theories have also been studied
extensively \cite{AS1}-\cite{Martin:2001ye} and two related 
features peculiar to noncommutative field theories
were found. They are ambiguities to define currents and IR singularity
in nonplanar diagrams.
If the fermions are in the fundamental representation, we can define
the following two different types of chiral currents in noncommutative
flat space, corresponding to different definitions of local chiral 
transformations:
\begin{eqnarray}
&& {\cal J}_{\mu,5}(x)=-i \psi_{\beta}(x)\star\bar
{\psi}_{\alpha}(x)(\gamma_{\mu}\gamma_5)^{\alpha \beta}, \n
&& {\cal J}'_{\mu,5}(x)=i \bar{\psi}_{\alpha}(x)\star
\psi_{\beta}(x)(\gamma_{\mu}\gamma_5)^{\alpha \beta}.
\end{eqnarray}
Since the fermions transform as $\psi \rightarrow U\psi$ under gauge
transformation, the first current transforms covariantly 
${\cal J}_{\mu} \rightarrow U{\cal J}_{\mu,5}U^{\dagger}$ 
while the second one invariantly. 

The covariant current is shown to satisfy a natural 
noncommutative generalization of the anomalous 
Ward--Takahashi(WT) identity (in d=4) 
\begin{equation}
\langle D^{\mu}{\cal J}_{\mu,5} \rangle \propto 
\epsilon^{\mu \nu \lambda \sigma} 
F_{\mu \nu} \star F_{\lambda \sigma},
\end{equation}
where an ordinary product is replaced by a noncommutative 
star product, and $D_{\mu}M=\partial_{\mu}M-ig[a_{\mu},M]$ 
is the covariant derivative for
adjoint representations.
It can be calculated diagrammatically, by point splitting
regularization, or by Fujikawa's method in path integrals.
Only planar diagrams contribute in the diagrammatic calculation to 
the anomaly in the r.h.s.
Both sides in the above equation transform as adjoint representations
under gauge transformation.

On the contrary, the gauge invariant current ${\cal J}'_{\mu}$ is 
invariant under gauge
transformation and the l.h.s. of the WT identity should be
$\partial^{\mu} {\cal J}'_{\mu,5}$ which is also gauge invariant.
Since there are no local gauge invariant quantities constructed from 
gauge fields in noncommutative field theories,
the r.h.s. cannot be expressed locally\cite{Banerjee:2001un}.
The authors in ref.\cite{AS2} obtained  a  nonlocal expression 
for the r.h.s. by calculating nonplanar diagrams.
It vanishes if we fix the external momenta finite ($k \theta \neq 0$)
and set the regularization cutoff $\Lambda$ infinity.
This result was  confirmed in refs.\cite{Armoni, Nakajima}.
On the other hand, if the external momenta is taken  to zero
before the cutoff $\Lambda$ is taken to infinity, a finite
anomaly term arises due to IR singularity.
The final result can be written by using a 
generalized star product, $\star'$\cite{AS2}. 

As we mentioned at the beginning, a noncommutative field theory
can be formulated as a matrix model and if we consider
a compact space, the system can be formulated in terms of 
finite dimensional matrices. One of the  simplest examples is 
a noncommutative
2-sphere (fuzzy 2-sphere). 
Since wave functions on fuzzy 2-sphere can be expanded in terms of 
noncommutative analogs of the spherical harmonics, which are 
constructed
from a finite dimensional representation of $SU(2)$ algebra,
field theory on it is formulated as a matrix model of finite
size.

In this paper, we consider  fermions in the fundamental representation
for gauge group on the above mentioned fuzzy 2-sphere 
and investigate their chiral properties, in particular,
calculate chiral anomaly. 
Chiral anomaly on the fuzzy 2-sphere has been examined 
in papers \cite{13_1,13_2,13_3} from various different approaches.
In ref.\cite{13_1}, chiral anomaly was discussed as a Jacobian 
term of the fermion measure under a chiral transformation. 
In ref.\cite{13_2}, it was discussed, along with topologically 
nontrivial field configurations, based on algebraic K-theory.
The model considered in ref.\cite{13_3} is similar to ours but 
their treatment is not completely satisfactory. 
In this paper, we make every step of calculations well-defined 
without introducing any regularizations or 
approximations, and clarify the distinction between the WT 
identities for a covariant and an invariant current. 
Our results are summarized in eq.(\ref{WIinvfinal}) 
and eq.(\ref{covWI2}).

In the system we consider, everything is finite dimensional and
no regularization is necessary. 
If both of the action and the measure were invariant under a
noncommutative chiral transformation on
fuzzy 2-sphere, we could not obtain the correct anomalous
WT identity for chiral currents. However, as we show in this paper, 
a Dirac operator on fuzzy 2-sphere is no longer anti-commutable 
with a chirality operator at finite $N$ ($N$ is the dimension of
matrices) and
a careful treatment of the $1/N$ correction will lead to
the correct form of anomalous WT identity for chiral
currents. In this sense, the chiral anomaly arises 
in a similar way to the case of the Wilson fermion 
in the lattice gauge theory\cite{KS}. 
In the case of Wilson fermion, the Wilson term which is 
introduced in the action
to remove  doublers violates chiral invariance of the action.
On the contrary, a  natural Dirac operator on fuzzy 2-sphere 
does not have the doubling problem and we do not need to add an 
extra term in the action to remove  doublers.
Nevertheless, an anti-commutator of the Dirac operator and 
a natural generalization of the chirality operator acquires an 
$1/N$ correction and this leads to the correct chiral anomaly.
We know the no-go theorem about chiral-invariant
Dirac operators in lattice gauge theories\cite{nielsen}.
We hope to have some analogous no-go theorem 
in theories on noncommutative space as well.

Like the flat noncommutative space, there are ambiguities for 
definitions of chiral currents, 
depending on definitions of
chiral transformations.
We define a gauge-invariant current and a gauge-covariant current.
The covariant current is shown to satisfy a local anomalous 
WT identity (\ref{covWI2})
while the invariant current satisfies  nonlocal one 
(\ref{WIinvfinal}).
These results are consistent with the previous results on
anomaly in flat space.

The organization of the paper is as follows.
In section 2, we explain briefly how to formulate 
noncommutative gauge theory on fuzzy 2-sphere as a matrix 
model of finite size.
We also show the spectrum of free Dirac operator and its
eigenfunctions. 
In section 3, we introduce two different limits, a commutative 
limit and
a flat limit.
Section 4 is the main part of the  paper. 
There we define
two different types of chiral transformations and calculate
anomalous WT identities for the corresponding chiral currents.
We calculate anomaly for the gauge invariant
current in subsection 4.1, 
and then for the gauge covariant current in subsection 4.2. 
Using various useful identities in appendices, we can 
calculate  anomalies for both cases.
Their local forms look very different.
We also consider a commutative and  flat limit.
Section 5 is devoted to discussions.
In appendices 6.1 and 6.2,
we summarize various useful identities 
which are used in the calculation of the WT identity.
In appendix 6.3, we review the calculation of anomaly
for the theory on the commutative 2-sphere in order to 
compare with the noncommutative result.
In appendix 6.4 some detailed calculations are shown.
\section{Dirac Operator on Fuzzy 2-Sphere}
\setcounter{equation}{0}
\subsection{Matrix Construction of Fuzzy 2-Sphere}
In order to define noncommutative geometry, Connes proposed to
generalize ordinary commutative wave functions 
to noncommutative ones, 
instead of making the geometry itself noncommutative.
In the case of noncommutative (fuzzy) 2-sphere, 
we first make the coordinates
noncommutative by using a
$(2L+1)$-dimensional representation of the angular momentum
operators $L_i$ which satisfy 
$[L_i,L_j]=i\epsilon_{ijk}L_k $ and 
$(L_i)^2=L(L+1)$.
We then introduce noncommutative coordinates on fuzzy 2-sphere by
\begin{equation}
x_i=\alpha L_i.
\end{equation}
They are noncommutative as
\begin{equation}
[x_i,x_j]=i\alpha \epsilon_{ijk}x_k,
\end{equation}
and form a sphere with a radius $\rho$ as,
\begin{equation}
(x_i)^2 =\alpha^2L(L+1)=\rho^2.
\end{equation}
Thus, $\alpha$ gives the noncommutative scale on fuzzy 2-sphere,
and the radius is related
to $\alpha$ and $L$ by the relation
$\rho=\sqrt{\alpha^2 L(L+1)}$.

Wave functions on fuzzy 2-sphere can be expanded in terms of noncommutative
analogs of the spherical harmonics which can be constructed from 
the above noncommutative coordinates.
Similarly to the commutative spherical harmonics,
products of $x_i$
can be decomposed into irreducible representations of $SO(3)$.
They are traceless symmetric products of $x_i$. 
Since the noncommutative coordinates $x_i$ are $(2L+1)$-dimensional
matrices, the noncommutative spherical functions are also 
matrices of the same size.
In the fuzzy sphere, there is an upper bound for the angular momentum
$l$ of the noncommutative spherical harmonics $\hat{Y}_{l,m}$; $l \le 2L$.
Any hermitian matrix $M$ can be expanded in terms of $\hat{Y}_{l,m}$ as
\begin{equation}
 M = \sum_{l=0}^{2L} \sum_{m=-l}^{l} m_{l,m} \hat{Y}_{l,m}.
\end{equation}
The total number of the basis wave functions 
is $\sum_{l=0}^{2L} (2l+1) =(2L+1)^2$ and
gives the number of independent basis of $(2L+1)$-dimensional hermitian matrices.
Some basic properties of $\hat{Y}_{l,m}$ are summarized in appendix 6.1.
In the following, we omit the hat  from $\hat{Y}_{l,m}$ unless there is any
confusion with the commutative spherical harmonics.

A noncommutative field theory on the fuzzy sphere can be formulated as
a matrix model by expanding fields in terms of the noncommutative spherical
harmonics as above. A hermitian matrix $M$ is mapped to an ordinary 
function on 2-sphere with the same coefficient $m_{l,m}$ as
\begin{equation}
 M \leftrightarrow M(\Omega)=\sum_{l=0}^{2L} \sum_{m=-l}^{l} m_{l,m} Y_{l,m}(\Omega),
\end{equation}
where $Y_{l,m}(\Omega)$ are ordinary spherical harmonics. 
Due to the noncommutativity of the coordinates, a product of two matrices is
mapped to the so-called star-product of functions on 2-sphere.
Derivatives on the fuzzy 2-sphere are expressed by taking a commutator with
the $SU(2)$ generator 
\begin{equation}
 {\cal L}_i M= [L_i, M] =(L_i^L-L_i^R)M \leftrightarrow \tilde{{\cal L}}_i M(\Omega) 
=-i \epsilon_{ijk} x_j \partial_{k} M(\Omega).
\label{adjointL}
\end{equation}
Here we have introduced a notation $L_i^L$ and $L_i^R$. The superscript $L$ or $R$
means that the operator acts on matrices by left or right
multiplication.
The superscript $L$ is often omitted in the following when there is no
confusion.
An integral over 2-sphere is replaced by taking trace over matrices
\begin{equation}
\frac{1}{2L+1} {\text Tr}\leftrightarrow \int 
\frac{d \Omega}{4 \pi}\equiv \int_\Omega .
\label{trintegral}
\end{equation}
More detailed relation between noncommutative field theories on the fuzzy 2-sphere and
matrix models is given in ref.\cite{IKTW}.
\subsection{Dirac Operator}
An action for  Dirac fermion on fuzzy 2-sphere is given by
\begin{eqnarray}
S_{S^2_F}&=& \frac{\alpha}{2g^2}{\text Tr}(\bar{\psi} D \psi)
, \label{FuzzyAction}\\
D&=&\sigma_i({\cal L}_i+\rho a_i)+1,  \label{Diracaction}
\end{eqnarray}
where $g$ is a coupling constant.
The spinor field $\psi$ and  the gauge field $a_i$ are 
$(2L+1)\times(2L+1)$ Hermitian matrices. 
Dirac fermions on the fuzzy 2-sphere were investigated  in 
refs.\cite{Grosse,Watamura}.
\par
This action (\ref{FuzzyAction}) is invariant under the 
following gauge transformation:
\begin{eqnarray}
\psi &\rightarrow& U\psi, \label{gaugeTrpsi}\\
\bar\psi &\rightarrow& \bar\psi U^\dag,\label{gaugeTrpsibar} \\
a_i &\rightarrow& U a_i U^\dag +\frac{1}{\rho} (U L_i U^\dag-L_i).
\label{gaugeTrai}
\end{eqnarray}
The last transformation is obtained from a requirement that
the combination
\begin{equation}
A_i=\alpha(L_i+\rho a_i)
\end{equation}
transforms covariantly under gauge transformation as
$A_i \rightarrow U A_i U^\dagger$.
The fermion $\psi$ transforms as  the fundamental 
representation.
The covariant derivative for  $\psi$ is given by
\begin{equation}
A_i' \psi = \alpha({\cal L}_i +\rho a_i) \psi.
\end{equation}
It is straightforward to see that $A_i' \psi$ transforms 
correctly as
\begin{equation}
A_i' \psi \rightarrow U A_i' \psi.
\end{equation}
\par
By noting that
$
\sigma_i {\cal L}_i =J_i^2-{\cal L}_i^2-\frac{3}{4} 
$
where
$ J_i={\cal L}_i +\frac{\sigma_i}{2}, $
 eigenvalues for the free Dirac operator
can be easily obtained:
\begin{eqnarray}
D_0&=&(\sigma_i {\cal L}_i +1) \label{freeDO} \\
&=&j(j+1)-l(l+1)+\frac{1}{4}\\
&=&\left\{
\begin{array}{lll}
l+1&=j+\frac{1}{2} &
{\rm for} \quad j=l+\frac{1}{2} \\
-l&-(j+\frac{1}{2}) &
{\rm for} \quad j=l-\frac{1}{2}. 
\end{array}
\right.
\end{eqnarray}
No doubler modes exist in the spectrum.  Various properties
of the Dirac operator on the fuzzy 2-sphere are discussed in
ref.\cite{bala}. 
\par
The eigenfunctions are given by spinorial-spherical-harmonics:
\begin{eqnarray}
{\cal Y}_{l+\frac{1}{2},m}&=&|j=l+\frac{1}{2},j_z=m \rangle 
\nonumber \\
&=&
\sqrt{\frac{l+\frac{1}{2}+m}{2l+1}} Y_{l,m-\frac{1}{2}}
\otimes |\uparrow \rangle
+
\sqrt{\frac{l+\frac{1}{2}-m}{2l+1}} Y_{l,m+\frac{1}{2}}
\otimes | \downarrow \rangle, \\
{\cal Y'}_{l-\frac{1}{2},m}&=&|j=l-\frac{1}{2},j_z=m \rangle 
\nonumber \\
&=&
\sqrt{\frac{l+\frac{1}{2}-m}{2l+1}} Y_{l,m-\frac{1}{2}}
\otimes |\uparrow \rangle
-
\sqrt{\frac{l+\frac{1}{2}+m}{2l+1}} Y_{l,m+\frac{1}{2}}
\otimes | \downarrow \rangle, 
\label{spinY}
\end{eqnarray}
which satisfy
\begin{eqnarray}
D_0{\cal Y}_{l+\frac{1}{2},m}&=&
(l+1){\cal Y}_{l+\frac{1}{2},m},\\
D_0{\cal Y'}_{l-\frac{1}{2},m}&=&
-l{\cal Y'}_{l-\frac{1}{2},m}.
\end{eqnarray}
Angular momentum $l$ takes values $l=0,1,\cdots ,2L$ for 
${\cal Y}_{l+\frac{1}{2},m}$ and $l=1,\cdots ,2L$ for
${\cal Y}'_{l-\frac{1}{2},m}$. Hence the eigenfunctions of
$D_0$ are paired between ${\cal Y}_{l+\frac{1}{2},m}$ and 
${\cal Y}'_{l+\frac{1}{2},m}$ 
with positive and negative eigenvalues
except for ${\cal Y}_{2L+\frac{1}{2},m}$.
\par
When we calculate anomalous WT identities for chiral
currents, we need to evaluate 
the following type of expectation values in the free Dirac 
action $S_0$:
\begin{eqnarray}
\langle O \rangle_{S_0}&=&\frac{1}{Z_{S_0}}
\int d\psi d\bar\psi O e^{-S_0},\\
Z_{S_0}&=&\int d\psi d\bar\psi e^{-S_0}, \ \ S_0= \frac{\alpha}{2g^2}{\text Tr}(\bar{\psi} D_0 \psi).
\end{eqnarray}
They can be calculated 
by expanding the fields $\psi$ and $\bar\psi$
in terms of 
 ${\cal Y}_{l+\frac{1}{2},m}$ and ${\cal Y'}_{l-\frac{1}{2},m}$, 
and then using  the Wick's theorem.
Expectation values for typical $O$'s are given in appendix 6.2.

\section{Commutative Limit and Flat Limit}
\setcounter{equation}{0}
In this section,
we will consider two different limits of the
action (\ref{FuzzyAction}), a commutative limit and a flat limit,
and obtain actions for  Dirac fermions 
on a commutative 2-sphere and a noncommutative 
flat-space\cite{stein}.
Then we will review chiral anomalies in these theory.
In the next section we will calculate the chiral anomaly
on the fuzzy 2-sphere, take these two limits, 
and then 
compare with the results of chiral anomalies that will have 
been reviewed in this section.
The calculation of the chiral anomaly in the next section 
can be done without introducing any kind of 
regularization and it is a nontrivial check
whether they agree or not.

\subsection{Commutative 2-Sphere}
For the commutative limit, we take the noncommutative
parameter $\alpha\to 0$, and the size of matrices $L\to \infty$,
with the radius $\rho$ fixed.
The action (\ref{FuzzyAction}) can be mapped 
to a noncommutative field
theory on the 2-sphere by using the mapping rules,
(\ref{adjointL}) and (\ref{trintegral}). Products are mapped to 
noncommutative star products.
In the commutative limit, this product can be approximated
by an ordinary product and we obtain
an action for the Dirac fermion on the commutative 2-sphere with 
the radius $\rho$:
\begin{eqnarray}
S_{S^2}&=&\frac{\rho}{g^2}\int_{\Omega}\bar\psi D\psi, \label{comAction}\\
D&=&D_0+\rho \sigma_i a_i, \\
D_0&=&\sigma_i \tilde{\cal L}_i +1.  
\end{eqnarray}
Here, the 2-sphere is embeded in a three-dimensional space,
and the Dirac operator can be rewritten as 
\begin{equation}
D=\rho[\sigma'_i(i\partial_i+a'_i)+\phi\gamma_3]+1, 
\end{equation}
where we have redefined new $\sigma$ matrices and new gauge fields for
later convenience:
\begin{eqnarray}
a'_i&=&\frac{1}{\rho}\epsilon_{ijk}x_j a_k, \label{aprime}\\
\sigma'_i&=&\frac{1}{\rho}\epsilon_{ijk}x_j \sigma_k. \label{sigmaprime}
\end{eqnarray}
They are tangential components of $a_i$ and $\sigma_i$. 
Similarly the normal components are given by
\begin{eqnarray}
\phi&=&\frac{1}{\rho}x_i a_i, \label{scalarphi}\\
\gamma_3&=&\frac{1}{\rho}x_i \sigma_i.
 \label{comgamma3}
\end{eqnarray}
They are a scalar field and  a chirality operator on 2-sphere, respectively. 
Hence the action (\ref{comAction}) contains
a scalar field $\phi$ as  the normal component of $a_i$. 
\par
We can further take a flat limit of (\ref{comAction}),
by considering the vicinity of the north pole on the 2-sphere,
taking $\rho \rightarrow \infty$, and mapping 
$\rho^2 \int _\Omega \rightarrow \int\frac{d^2 x}{4\pi}$.
Then we obtain 
\begin{equation}
S_{R^2}=\frac{1}{4\pi g^2}\int d^2 x \bar\psi
[\sigma''_i(i\partial_i+a''_i)+\phi\sigma_3]\psi,
\label{comflatAction}
\end{equation}
with
\begin{eqnarray}
\sigma''_i&=&-\epsilon_{ij}\sigma_j,  \label{sigmapp}\\
a''_i&=&-\epsilon_{ij}a_j,  \label{app}
\end{eqnarray}
where $i,j = 1,2$.
This is nothing but an action for Dirac fermions 
on  a 2-dimensional commutative flat-space with a 
Yukawa coupling to a scalar field.
\par
A chiral transformation is defined on the commutative 2-sphere as
\begin{eqnarray}
\delta \psi &=&\lambda\gamma_3 \psi,\label{comCT}\\
\delta \bar\psi &=& \lambda \bar\psi \gamma_3.
\end{eqnarray}
where the chirality operator of (\ref{comgamma3}) satisfies
\begin{eqnarray} 
(\gamma_3)^\dagger&=&\gamma_3,\\
(\gamma_3)^2 &=&1,\\ 
\left\{D,\gamma_3 \right\}&=&2\rho\phi.
\end{eqnarray}

Anomalous chiral WT identity for the theory
on the commutative 2-sphere has been calculated in 
ref.\cite{wong}. Since now we consider the theory with 
the Yukawa coupling to the scalar filed $\phi$, 
we show the calculation for the chiral WT identity
in the appendix \ref{sec:comWI}.
The result is
\begin{equation}
\tilde{\cal L}_i (\bar\psi \sigma_i \gamma_3 \psi) 
-2\rho\phi\bar\psi \psi 
= 
\frac{g^2}{\rho}(-4i\epsilon_{ijk} x_k (\tilde{\cal L}_i a_j) 
-8\rho\phi).\label{comAWI1}
\end{equation}
This can be rewritten as a more familiar form
\begin{equation}
i\partial_i (\bar\psi \sigma'_i \gamma_3 \psi) 
-2\phi\bar\psi \psi 
= 
\frac{2g^2}{\rho}\epsilon_{ijk} x_i (\partial_j a'_k-\partial_k a'_j),
\label{comAWI2}
\end{equation}
by separating $a_i$ and $\sigma_i$ into tangential and normal
components as in
 eq.(\ref{aprime}),(\ref{sigmaprime}),(\ref{scalarphi}), (\ref{comgamma3}).
\par
Taking the flat limit further, we obtain
\begin{equation}
i\partial_i (\bar\psi \sigma''_i \gamma_3 \psi) 
-2\phi\bar\psi \psi 
= 
2g^2 \epsilon_{ij}(\partial_i a''_j-\partial_j a''_i),
\label{comflatWI}
\end{equation}
which agrees with the WT identity in 2-dimensional flat-space.
Note that in  a usual convention the r.h.s. is divided by $4\pi$.
This can be given by multiplying the action (\ref{comAction}),
and then (\ref{comflatAction}), by $4\pi$.

\subsection{Noncommutative Flat Space}
Now we take the flat limit from the fuzzy 2-sphere
with the noncommutativity fixed.
This can be done 
by considering the vicinity of the north pole on the fuzzy 2-sphere,
and taking the radius $\rho \rightarrow \infty$,
the system size $L \rightarrow \infty$,
with the noncommutative parameter $\theta=\alpha\rho$ fixed.
In this limit, a matrix M is mapped to a function on the 
flat space and  ${\text Tr}$ becomes an integral over this flat space:
\begin{equation}
\frac{1}{2L+1}{\text Tr}
\rightarrow
\frac{1}{4\pi\rho^2}\int dx^2.
\label{flatint}
\end{equation}
In the vicinity of the north pole, $x_3$ can be replaced by
$\rho$ and a commutator between $x_i$ ($i=1,2$) becomes
$[x_i,x_j]=i\theta\epsilon_{ij}$. Hence 
\begin{eqnarray}
[L_i,M]
&=&\frac{1}{\alpha}[x_i,M] \n
&\rightarrow& i\frac{\theta}{\alpha}\epsilon_{ij}\partial_{j}M(x)
=i\rho\epsilon_{ij}\partial_{j}M(x).
\label{flatadjLi}
\end{eqnarray}
Then, from the action on the fuzzy 2-sphere (\ref{FuzzyAction}),
we obtain an action for the Dirac fermion 
on the noncommutative 2-dimensional flat-space:
\begin{equation}
S_{R^2_{\rm NC}}
=\frac{1}{4\pi g^2}\int d^2 x
\bigl[\bar\psi[\sigma''_i (i\partial_i +a''_i)
+\sigma_3\phi]\psi\bigr]_\star,
\label{flatAction}
\end{equation}
where  $\sigma''_i$ and $a''_i$ are defined in eq.(\ref{sigmapp}),
(\ref{app}),
and $[\cdots]_\star$ means that any product in the bracket 
is considered as a star product. 

By taking  $\theta \rightarrow 0$,
we can obtain the commutative limit of
(\ref{flatAction}), which again becomes
(\ref{comflatAction}).
In the previous subsection,
we first took the commutative limit, 
$\alpha \rightarrow 0$ with $\rho$ fixed,
and then took the flat limit $\rho \rightarrow \infty$,
while here we first took the flat limit,
$\rho \rightarrow \infty$ with $\theta=\alpha\rho$ fixed,
and then the commutative limit, $\theta \rightarrow 0$.
We arrived at the same classical action
irrespective of the ordering of 
taking limits, though there may be some subtle 
dynamics quantum mechanically. 
\par
As we have reviewed in the introduction,
chiral anomaly on noncommutative flat-space has been investigated 
in a number of papers \cite{AS1}-\cite{Martin:2001ye}.

\section{Chiral Anomaly on Fuzzy 2-Sphere}
In this section, we define chiral transformations and calculate 
the chiral anomaly for fermions in the fundamental representation
on the  fuzzy 2-sphere (\ref{FuzzyAction}).
If we keep the ordering of 
$\bar{\psi}$ and $\psi$ in (\ref{comAWI1}),
the l.h.s is gauge invariant and 
it will be difficult to write down its noncommutative analog
since we cannot make gauge invariant local operators on 
noncommutative space from gauge fields.
We have to take trace over matrices
to make an operator gauge invariant.
Hence, the WT identity with a gauge-invariant current cannot 
be written down unless we introduce explicit nonlocality.
On the other hand, local WT identity can be written down if we use a
gauge-covariant current, instead. 
They have been discussed fully in the case of flat 
noncommutative space
as  reviewed in the introduction.
In the following,  we will consider two kinds of chiral
transformations and the corresponding chiral currents:
in a gauge invariant way and a covariant way. 

\subsection{Gauge-Invariant Current}
We define a local chiral transformation as
\begin{eqnarray}
\delta\psi&=&\frac{1}{2L+1}\left(\sigma_i \psi 
\left\{\lambda, L_i \right\} -\psi\lambda \right),\label{invCT} \\
\delta\bar\psi&=&\frac{1}{2L+1}
\left( \left\{\lambda , L_i \right\}
\bar\psi \sigma_i -\lambda \bar\psi \right),
\end{eqnarray}
where the transformation parameter $\lambda$ is also 
a $(2L+1)\times (2L+1)$ matrix and anti-hermitian.
This chiral transformation reduces to the ordinary one 
(\ref{comCT}) in the commutative limit.
Furthermore, a chiral operator 
\begin{equation}
\Gamma^R=\frac{1}{2L+1}(2\sigma_i L_i^R -1),
\label{GammaR}
\end{equation}
satisfies
\begin{eqnarray}
(\Gamma^R)^2&=&1,\\
(\Gamma^R)^\dagger&=&\Gamma^R.
\end{eqnarray}
at finite $L$. Various properties of this chiral operator 
are discussed in ref.\cite{bala}.
The chiral operator (\ref{GammaR}) 
reduces to the ordinary chiral operator (\ref{comgamma3})
in the commutative limit. 
An anticommutator with the Dirac operator 
(\ref{Diracaction}) becomes
\begin{equation}
\{\Gamma^R,D \}= \frac{1}{L+1/2} \left(
2({\cal L}_i+\rho a_i)L_i^R - D
\right). \label{gammaDac}
\end{equation}
Here ${\cal L}_i L_i^R$ and $D/(L+1/2)$ vanish in the 
commutative limit since they are of order $1/L$.
Hence the anti-commutator becomes proportional to the 
scalar field as expected:
\begin{equation}
\{\Gamma^R,D \} \rightarrow 2  a_i x_i = 2 \rho \phi.
\end{equation}
The r.h.s. of the anticommutator (\ref{gammaDac}) has correction terms of order $1/L$.
These terms lead to the correct anomaly as we show in the following.

%
\par
The chiral transformation (\ref{invCT})
is compatible with the gauge transformation if
 $\lambda$ is a gauge invariant parameter.
Namely, $\delta \psi$ also transforms as the fundamental representation
$\delta \psi \rightarrow U \delta \psi $.
This is because $\lambda$ and $L_i$ are placed in  the right of
$\psi$ in (\ref{invCT}).
Since the chiral transformation parameter $\lambda$
is invariant under gauge transformation,
the chiral current associated with this transformation is 
also gauge invariant.

\subsubsection{Evaluation of the WT Identity}
Now we will evaluate the WT identity associated with the 
chiral transformation (\ref{invCT}).
Under the chiral transformation, 
the variation of 
the action (\ref{FuzzyAction})
can be obtained by using the anticommutator (\ref{gammaDac}) as
\begin{eqnarray}
\delta S_{S^2_F} 
&=& \frac{\alpha}{g^2 (2L+1)}{\text Tr}
\Bigl[\frac{1}{2}[L_i,\lambda]
\left(
\left\{L_j,\bar\psi\sigma_i\sigma_j\psi \right\} 
-\bar\psi\sigma_i \psi \right) \n
&& 
+\left\{\lambda,L_i \right\}
\left(\bar\psi [L_i,\psi]
+\rho \bar\psi a_i \psi\right) 
-\lambda \bar\psi D \psi
\Bigr]. \label{deltaS2}
\end{eqnarray}
The integration measure also varies as
$d\psi' d\bar\psi'= J d\psi d\bar\psi$ 
with the Jacobian,
\begin{equation}
J=
\left|
\begin{array}{cc}
\frac{\partial \psi'}{\partial \psi} &
\frac{\partial \psi'}{\partial \bar\psi} \\
\frac{\partial \bar\psi'}{\partial \psi} &
\frac{\partial \bar\psi'}{\partial \bar\psi} 
\end{array}
\right|^{-1} 
=1+4{\text Tr}\lambda+{\cal O}(\lambda^2).
\end{equation}
By combining both variations under the chiral transformation,
we have the anomalous chiral WT identity:
\begin{equation}
 \langle \delta S_{S^2_F}\rangle_{S}  -4{\text Tr}\lambda =0. 
\label{Wid}
\end{equation}
The contribution from the Jacobian and the last term in
(\ref{deltaS2}) are cancelled by the relation 
\footnote{If we define chiral transformation for $\psi$ without the
second term in (\ref{invCT}), 
either of the last term in (\ref{deltaS2}) 
or the contribution from the Jacobian 
do not appear from the beginning.
This chiral transformation also agrees with the 
commutative one in the 
commutative limit, 
and there is no reason to exclude this simpler 
transformation. But the resulting WT identity becomes the same
except for a slight change (of order $1/L$) of the definition of the chiral current.
This property is desirable since ambiguities in making noncommutative
systems  from a commutative one do not affect universal
structures of noncommutative systems such as anomaly.
}
\begin{equation}
\frac{\alpha}{g^2 (2L+1)}
\langle {\text Tr} \left[ \lambda
\bar\psi D\psi \right]
\rangle_S 
=-4{\text Tr}\lambda.
\end{equation}
This can be easily proved by formally diagonalizing the operator $D$.
\par
The first term in (\ref{deltaS2}) proportional to $[L_i,\lambda]$ gives
the divergence of the chiral current in the WT identity.
The second term can be rewritten as 
\begin{eqnarray}
&& Tr \left( \left\{\lambda,L_i \right\}
\left(\bar\psi [L_i,\psi]
+\rho \bar\psi a_i \psi\right) \right) \n
&& =Tr \left(
 [\lambda, L_i]\left( \bar\psi [L_i,\psi]
+\rho \bar\psi a_i \psi\right)  + 2 L_i \lambda \bar\psi[L_i, \psi]
+ 2 \rho \lambda \bar\psi a_i \psi L_i  
\right).
\label{WIcal}
\end{eqnarray}
The first term in (\ref{WIcal}) is absorbed in the definition
of the chiral current. In the following analysis, we
consider terms up to the first order in the gauge fields $a_i$.
We make use of the identity (see eq.(\ref{psiApsiBs0}))
\begin{equation}
\langle {\text Tr}(\lambda \bar\psi a_i 
[\psi,L_i])\rangle_{S_0} =0,
\end{equation}
and define the scalar field,
\begin{equation}
\phi=\frac{\alpha}{2\rho}
\left[\{a_i,L_i\}+\rho a_i^2 \right]
=\frac{\alpha}{2\rho^2}
\left[ (L_i+\rho a_i)^2-(L_i)^2\right],
\label{scalarfield}
\end{equation}
which is a normal 
component of $a_i$, and transforms 
covariantly under the gauge transformation.
Then, we can write down the WT identity as
\begin{eqnarray}
&&\langle {\text Tr}\left(\frac{\lambda}{2}
\left[L_i,\left\{L_j,\bar\psi\sigma_i\sigma_j\psi \right\} 
-\bar\psi\sigma_i\psi 
-2\bar\psi[L_i,\psi]-2\rho\bar\psi a_i \psi
\right]
\right) 
-\frac{2\rho^2}{\alpha}
{\text Tr}\left(\lambda \bar\psi \phi\psi\right)
\rangle_S \n
&=&
-\rho\langle{\text Tr}\bar\psi [L_i,a_i]\psi\rangle_{S_0} 
-\langle {\text Tr}\left( 2L_i \lambda\bar\psi [L_i,\psi]
\right)
{\text Tr}\left(\frac{\alpha\rho}{2g^2}
\bar\psi \sigma_i a_i \psi \right)
\rangle_{S_0} 
+{\cal O}((a_i)^2). \n
&&
\label{invWI1}
\end{eqnarray}
The last term in (\ref{invWI1}) came from
the first-order perturbative expansion with respect to
the gauge field in the action (\ref{FuzzyAction}).
\par
Now we evaluate the last term in eq.(\ref{invWI1}).
Here we assume that the background gauge field has only the third 
component $a_3$. We can recover the complete result afterward
by using $SO(3)$ invariance.
By using the formula (\ref{psiApsiBpsiCsigma3psiDs0}), we have
\begin{eqnarray}
&&-\langle {\text Tr}
\left(2\bar\psi \left[L_i,\psi\right]L_i \lambda\right)
{\text Tr}\left(\frac{\alpha\rho}{2g^2}
\bar\psi\sigma_3 a_3 \psi \right)
\rangle_{S_0} \n
&=&
-\frac{2g^2\rho}{\alpha (2L+1)^2}
\sum_{l=1}^{2L}\sum_{l'=1}^{2L}
\sum_{m=-l+\frac{1}{2}}^{l-\frac{1}{2}}
\sum_{m=-l'+\frac{1}{2}}^{l'-\frac{1}{2}}
\frac{1}{l(l+1)}\sqrt{\left(l+\frac{1}{2}\right)^2-m^2}
\sqrt{\left(l'+\frac{1}{2}\right)^2-{m'}^2} \nonumber \\
&&
\times\left[{\text Tr}(Y_{l,m+\frac{1}{2}}^\dag 
Y_{l',m'+\frac{1}{2}}\lambda)
{\text Tr}(Y_{l',m'-\frac{1}{2}}^\dag a_3
Y_{l,m-\frac{1}{2}}) \right. 
 \left.
-{\text Tr}(Y_{l,m-\frac{1}{2}}^\dag 
Y_{l',m'-\frac{1}{2}}\lambda)
{\text Tr}(Y_{l',m'+\frac{1}{2}}^\dag a_3
Y_{l,m+\frac{1}{2}})
\right] \n
&&
-\frac{4g^2\rho}{\alpha (2L+1)^2}
\sum_{l=1}^{2L}\sum_{m=-l}^{l}
m{\text Tr}(\lambda Y_{lm}){\text Tr}(Y_{lm}^\dag a_3).
\label{secondinvWI}
\end{eqnarray}
The last term of (\ref{secondinvWI}) is simplified
by using the completeness of the spherical harmonics (\ref{compYtt})
as
\begin{equation}
-\frac{4g^2\rho}{\alpha (2L+1)^2}
\sum_{l=1}^{2L}\sum_{m=-l}^{l}
{\text Tr}(\lambda Y_{lm})
{\text Tr}(Y_{lm}^\dag \left[L_3,a_3 \right])
=
-\frac{4g^2\rho}{\alpha (2L+1)}
{\text Tr}(\lambda \left[L_3,a_3 \right]),
\end{equation}
{}From $SO(3)$ invariance, it gives
\begin{equation}
-\frac{4g^2\rho}{\alpha (2L+1)}
{\text Tr}(\lambda \left[L_i,a_i \right]),
\end{equation}
which exactly cancels with the first term of 
the r.h.s. in 
eq.(\ref{invWI1}).
Hence the first term of (\ref{secondinvWI}) gives 
the r.h.s. of the WT identity (\ref{invWI1}).
\par
The first term of (\ref{secondinvWI}),
which we call $H_3$, can be evaluated as
\begin{eqnarray}
H_3&=&\frac{2g^2 \rho}{\alpha (2L+1)}\sum_{l=1}^{2L}
\sum_{l'=1}^{2L} \sum_{m=-l+\frac{1}{2}}^{l-\frac{1}{2}}
\sum_{m'=-l'+\frac{1}{2}}^{l'-\frac{1}{2}} \frac{1}{l(l+1)}
\nonumber \\
&&\times \left[
{\text Tr}(Y_{l,m+\frac{1}{2}}^\dag  
Y_{l',m'+\frac{1}{2}}\lambda)
{\text Tr}\left(
[L_+,Y_{l',m'+\frac{1}{2}}^\dag ] a_3 
[L_-,Y_{l,m+\frac{1}{2}}]  \right) \right. \nonumber \\
&&\left.
-{\text Tr}(Y_{l,m-\frac{1}{2}}^\dag 
Y_{l',m'-\frac{1}{2}}\lambda )
{\text Tr}\left(
[L_-,Y_{l',m'-\frac{1}{2}}^\dag ] a_3 
[L_+,Y_{l,m-\frac{1}{2}}]  \right)
\right] \nonumber \\
&=&\frac{2g^2 \rho}{\alpha (2L+1)}\sum_{l=1}^{2L}
\sum_{m=-l}^{l} \frac{1}{l(l+1)}
\nonumber \\
&&\times \left[
{\text Tr}\left(
[L_+,\lambda Y_{l,m}^\dag ] a_3 
[L_-,Y_{l,m}]  \right)
-{\text Tr}\left(
[L_-,\lambda Y_{l,m}^\dag ] a_3 
[L_+,Y_{l,m}]  \right)
\right]. \label{H31} \nonumber \\
\end{eqnarray}
Here we have used (\ref{L+Y}), (\ref{L-Y}), and then 
(\ref{compYtt}).
Using the identity (\ref{idll1}), and then (\ref{compYt}), 
this term becomes 
\begin{eqnarray}
H_3&=&
\frac{2g^2 \rho}{\alpha (2L+1)}\frac{1}{2L(L+1)}\sum_{l=1}^{2L}
\sum_{m=-l}^{l} \sum_{n=0}^\infty
\frac{1}{\{L(L+1)\}^n} 
 \nonumber \\
&&\times \left[
{\text Tr}\left(
[L_+,\lambda Y_{l,m}^\dag ] a_3 
\left[L_-,(L_i^L L_i^R)^n 
Y_{l,m}\right]  \right) \right. \nonumber 
%
\left.
-{\text Tr}\left(
[L_-,\lambda Y_{l,m}^\dag ] a_3 
\left[L_+,(L_i^L L_i^R )^n 
Y_{l,m}\right]  \right)
\right] \\
&=&
\frac{g^2\rho}{\alpha}\sum_{n=0}^{\infty}\frac{1}{\{L(L+1)\}^{n+1}} 
\Bigl[
{\text Tr}\left(L_{i_n}\cdots L_{i_1} L_+ \lambda \right)
{\text Tr}\left(a_3 L_-L_{i_1}\cdots L_{i_n} \right) \n
&&-{\text Tr}\left(L_{i_n}\cdots L_{i_1} L_- L_+ \lambda \right)
{\text Tr}\left(a_3 L_{i_1}\cdots L_{i_n} \right) 
-{\text Tr}\left(L_{i_n}\cdots L_{i_1} \lambda \right)
{\text Tr}\left(L_+ a_3 L_- L_{i_1}\cdots L_{i_n} \right)\n 
&&+{\text Tr}\left(L_{i_n}\cdots L_{i_1} L_- \lambda \right)
{\text Tr}\left(L_+ a_3 L_{i_1}\cdots L_{i_n} \right) 
-[L_+ \leftrightarrow L_-] 
\Bigr]. 
\end{eqnarray}
By $SO(3)$ invariance, the from in general
background gauge configurations can be determined as
\begin{eqnarray}
H
&=&
\frac{-2i\epsilon_{ijk}g^2\rho}{\alpha}
\sum_{n=0}^{\infty}\frac{1}{\{L(L+1)\}^{n+1}} \n
&&
\times\Bigl[
{\text Tr}\left(L_{i_n}\cdots L_{i_1} L_i \lambda \right)
{\text Tr}\left(a_k L_jL_{i_1}\cdots L_{i_n} \right) 
-{\text Tr}\left(L_{i_n}\cdots L_{i_1} L_j L_i \lambda \right)
{\text Tr}\left(a_k L_{i_1}\cdots L_{i_n} \right) \n
&&-{\text Tr}\left(L_{i_n}\cdots L_{i_1} \lambda \right)
{\text Tr}\left(L_i a_k L_j L_{i_1}\cdots L_{i_n} \right) 
+{\text Tr}\left(L_{i_n}\cdots L_{i_1} L_j \lambda \right)
{\text Tr}\left(L_i a_k L_{i_1}\cdots L_{i_n} \right) 
\Bigr] \n 
&=&
\frac{-2ig^2\rho}{\alpha}
\sum_{n=0}^{\infty}\frac{1}{\{L(L+1)\}^{n+1}} \n
&&\times\Bigl[
{\text Tr}\left(L_{i_n}\cdots L_{i_1} \lambda\right)
\bigl[\epsilon_{ijk}{\text Tr}\left(L_i[L_j,a_k]L_{i_1}\cdots L_{i_n} \right) 
-i{\text Tr}\left(L_i a_i L_{i_1}\cdots L_{i_n} \right)\bigr] \n
&&-{\text Tr}\left(L_{i_n}\cdots L_{i_1} L_i \lambda\right)
\bigl[\epsilon_{ijk}{\text Tr}\left([L_j,a_k]L_{i_1}\cdots L_{i_n} \right) 
-i{\text Tr}\left(a_i L_{i_1}\cdots L_{i_n} \right)\bigr]
\Bigr]. 
\label{invanoterm}
\end{eqnarray}
Finally, the WT identity (\ref{invWI1}) for the gauge invariant chiral current becomes
\begin{equation}
\langle {\text Tr}\left(\frac{\lambda}{2}
\left[L_i,J_i^5 \right]
\right) 
-\frac{2\rho^2}{\alpha}
{\text Tr}\left(\lambda \bar\psi \phi\psi\right)
\rangle_S 
=H,
\label{WIinvfinal}
\end{equation}
where the chiral current is given by
\begin{equation}
 J_i^5=\left\{L_j,\bar\psi\sigma_i\sigma_j\psi \right\} 
-\bar\psi\sigma_i\psi 
-2\bar\psi[L_i,\psi]-2\rho\bar\psi a_i \psi.
\end{equation}
The r.h.s. of the WT identity (\ref{WIinvfinal}), $H$, contains,
in addition to an anomaly term, 
an extra term which makes the scalar term in l.h.s.
of (\ref{WIinvfinal}) normal ordered.
We will see this in the next subsection.
Note  that we have considered terms up to the first order in
the gauge fields,
and there are higher order terms 
which are neglected in the above WT identity.
\par
Generic property of the WT identity for the gauge invariant current
is that the chiral transformation parameter $\lambda$ and 
the gauge field $a_i$ are inserted in  different traces in 
(\ref{invanoterm}).
This means that we cannot write down explicit local
expressions for the WT identity because the gauge field
$a_i$ is always in a trace, namely, in an integral 
over the sphere.
This is consistent with a general argument
that there is no local gauge invariant quantity
constructed from gauge fields on noncommutative space. 

\subsubsection{$\lambda =1$ Case}
In order to see that $H$ contains an extra term (vev of the scalar term)
mentioned above,
we evaluate $H$ 
for a special case $\lambda=1$.
We again set the background gauge configuration $a_i=a_3 \delta_{i3}$.
Then $H_3$ becomes
\begin{eqnarray}
H_3 |_{\lambda=1} 
&=& 
-\frac{2g^2\rho}{\alpha (2L+1)^2}\sum_{l=1}^{2L}
\sum_{m=-l+\frac{1}{2}}^{l-\frac{1}{2}}
\frac{1}{l(l+1)} 
\left\{ \left(l+\frac{1}{2}\right)^2 -m^2 \right\} \nonumber \\
&&
\times\left[ 
{\text Tr}(Y_{l,m-\frac{1}{2}}^\dag a_3 
Y_{l,m-\frac{1}{2}}) 
-{\text Tr}(Y_{l,m+\frac{1}{2}}^\dag a_3 Y_{l,m+\frac{1}{2}})
\right] \label{H3lambda11} \nonumber \\
&=&
\frac{4g^2\rho}{\alpha(2L+1)}\sum_{l=1}^{2L}\sum_{m=-l}^l
\frac{m}{l(l+1)} {\text Tr}(Y_{lm}^\dag a_3 Y_{lm}) . 
\label{H3lambda12}
\end{eqnarray}
The summation over $l,m$ in this equation can be evaluated
by using  the identity (\ref{idll1}) and 
the completeness of the spherical harmonics (\ref{compYt}).
We refer the detailed calculation to appendix  \ref{H3lambda1}.
We finally obtain
\begin{equation}
H_3 |_{\lambda=1}=\frac{8g^2\rho}{\alpha(2L+1)} 
{\text Tr}(a_3 L_3).\label{H3lambda13}
\end{equation} 
{}From $SO(3)$ invariance, a general form is given by
\begin{equation}
H|_{\lambda=1}=\frac{8g^2\rho}{\alpha(2L+1)} {\text Tr}(a_i L_i).
\label{Hlambda1}
\end{equation}
Because of the relation
\begin{equation}
 -\frac{2 \rho^2}{\alpha} \langle {\text Tr}  \bar\psi \phi \psi
\rangle_{S_0} = \frac{8g^2\rho}{\alpha(2L+1)} 
{\text Tr}(a_3 L_3),
\end{equation}
(\ref{Hlambda1}) becomes a vev of the scalar
term in the l.h.s. of WT identity (\ref{invWI1}),
and can be interpreted as a normal ordering constant for it.

\subsubsection{Flat Limit}
The flat noncommutative limit corresponds to considering a vicinity
of the north pole. 
More precisely, this means that the support of the background
gauge field $a_i$ and the chiral transformation parameter $\lambda$
are localizes around the north pole, and $L_i$ in the traces
including these fields can be replaced by $L_3$
in the leading order.
Hence the flat limit of the anomaly term 
(\ref{invanoterm}) vanishes because the second line 
and the third line of (\ref{invanoterm})
are cancelled each other. 
On the contrary, 
in an almost constant chiral transformation where 
$\lambda$ is close to 1, $L_i$ in the trace in which $\lambda$
is inserted cannot be replaced with $L_3$ and the cancellation
does not occur.
These results are
consistent with the calculations in the flat 
noncommutative space\cite{AS2, Armoni}.

\subsection{Gauge-Covariant Current}
We define another type of a chiral transformation as
\begin{eqnarray}
\delta\psi&=&\frac{1}{L+\frac{1}{2}}\sigma_i\lambda\psi L_i, 
\label{covCT}\\
\delta\bar\psi&=&\frac{1}{L+\frac{1}{2}}L_i\bar\psi\sigma_i\lambda.
\end{eqnarray}
This chiral transformation reduces to (\ref{comCT})
in the commutative limit.
The algebra of the chiral transformation closes 
up to the gauge transformation since
\begin{equation}
\delta_1\delta_2\psi=
\frac{1}{(L+\frac{1}{2})^2}\sigma_i\lambda_1\lambda_2\psi L_i
+\frac{L(L+1)}{(L+\frac{1}{2})^2}\lambda_1\lambda_2\psi.
\end{equation}
The chiral transformation (\ref{covCT}) is compatible with 
the gauge transformation, (\ref{gaugeTrpsi}),(\ref{gaugeTrpsibar})
if $\lambda$ transforms covariantly as
\begin{equation}
\lambda \rightarrow U \lambda U^\dag. \label{covlambdatr}
\end{equation}
Since the chiral transformation parameter $\lambda$ transforms
gauge covariantly as (\ref{covlambdatr}), the associated current
is also gauge covariant. Thus, the local WT identity is
expected to be written down.
\par 
\subsubsection{Evaluation of the WT Identity}
Now, we will evaluate the chiral WT identity for the covariant
current.
Under the chiral transformation (\ref{covCT}), 
the action (\ref{FuzzyAction}) varies as 
\begin{eqnarray}
\delta S_{S^2_F} &=&\frac{\alpha}{g^2(2L+1)}{\text Tr}
\Bigl(\bar\psi\sigma_i\sigma_j\left[L_i+\rho a_i, \lambda \right]\psi L_j 
+ 2\bar\psi\lambda\left[L_i,\psi \right] L_i \n
&& +\frac{2\rho^2}{\alpha}\bar\psi\lambda\phi\psi
-\rho^2 \bar\psi\lambda a_i^2\psi 
-2\rho\bar\psi \lambda a_i\left[L_i,\psi \right]
-\rho\bar\psi\lambda\left[L_i,a_i \right]\psi
\Bigr),
\end{eqnarray}
where
$\phi$ is the gauge covariant scalar field defined in 
(\ref{scalarfield}).
The integration measure is invariant.
Therefore, the WT identity becomes 
\begin{eqnarray}
&&
\langle {\text Tr}
\left(\lambda\left[L_i+\rho a_i, -\psi_\beta L_j\bar\psi_\alpha 
(\sigma_i \sigma_j)^{\alpha\beta} \right] \right)
-\frac{2\rho^2}{\alpha}
{\text Tr}\left(\bar\psi \lambda \phi\psi \right) 
\rangle_S \n
&=&
-\rho\langle {\text Tr}\left(\bar\psi \lambda\left[L_i,a_i \right]
\psi \right) \rangle_{S_0}\n
&&-\langle {\text Tr}
\left(2\bar\psi \lambda \left[L_i,\psi\right]L_i\right)
{\text Tr}\left(\frac{\alpha\rho}{2g^2}\bar\psi\sigma_j a_j \psi \right)
\rangle_{S_0} 
+{\cal O}((a_i)^2)
\label{covWI1}   
\end{eqnarray}
up to the first order in the gauge field $a_i$.
The last term came from 
the first-order perturbative expansion in the gauge fields
in (\ref{FuzzyAction}).
\par
Now we evaluate the last term in eq.(\ref{covWI1})
for a special background gauge field, where $\sigma_j a_j$
is replaced by $\sigma_3 a_3$.
By using the formula (\ref{psiApsiBpsiCsigma3psiDs0}), it can be
rewritten as
\begin{eqnarray}
&&-\langle {\text Tr}
\left(2\bar\psi \lambda \left[L_i,\psi\right]L_i\right)
{\text Tr}\left(\frac{\alpha\rho}{2g^2}
\bar\psi\sigma_3 a_3 \psi \right)
\rangle_{S_0} \n
&=&
-\frac{2g^2\rho}{\alpha (2L+1)^2}
\sum_{l=1}^{2L}\sum_{l'=1}^{2L}
\sum_{m=-l+\frac{1}{2}}^{l-\frac{1}{2}}
\sum_{m=-l'+\frac{1}{2}}^{l'-\frac{1}{2}}
\frac{1}{l(l+1)}\sqrt{\left(l+\frac{1}{2}\right)^2-m^2}
\sqrt{\left(l'+\frac{1}{2}\right)^2-{m'}^2} \nonumber \n
&&
\times\left[{\text Tr}(Y_{l,m+\frac{1}{2}}^\dag \lambda
Y_{l',m'+\frac{1}{2}})
{\text Tr}(Y_{l',m'-\frac{1}{2}}^\dag a_3
Y_{l,m-\frac{1}{2}}) \right. 
 \left.
-{\text Tr}(Y_{l,m-\frac{1}{2}}^\dag \lambda
Y_{l',m'-\frac{1}{2}})
{\text Tr}(Y_{l',m'+\frac{1}{2}}^\dag a_3
Y_{l,m+\frac{1}{2}})
\right] \n
&&
-\frac{4g^2\rho}{\alpha (2L+1)^2}
\sum_{l=1}^{2L}\sum_{m=-l}^{l}
m{\text Tr}(\lambda Y_{lm}){\text Tr}(Y_{lm}^\dag a_3).
\label{secondcovWI}
\end{eqnarray}
As in (\ref{secondinvWI}),
the last term of (\ref{secondcovWI}) can be calculated 
by using the completeness of the spherical harmonics (\ref{compYtt}),
and exactly cancels with the first term of 
the r.h.s. in eq.(\ref{covWI1}).
Hence the first term of (\ref{secondcovWI})
gives the r.h.s. of 
the WT identity (\ref{covWI1}).

The first term of (\ref{secondcovWI}), 
which we call $I_3$, 
when $\lambda=1$,
turns out to be exactly equal to $H_3|_{\lambda=1}$ 
(see eq.(\ref{H3lambda11})).
Following the same steps in $H_3$ case, we obtain
\begin{equation}
I|_{\lambda=1}=H|_{\lambda=1}=
\frac{8g^2\rho}{\alpha(2L+1)^2}{\text Tr}(a_i L_i).
\end{equation}
This term 
corresponds to the normal-ordering constant
for the scalar term in the l.h.s. of the WT identity
(\ref{covWI1}). 

We then evaluate the first term of (\ref{secondcovWI}) 
for $\lambda \neq 1$.
Using (\ref{L+Y}) and (\ref{L-Y}),
\begin{eqnarray}
I_3&=&\frac{2g^2 \rho}{\alpha (2L+1)}\sum_{l=1}^{2L}
\sum_{l'=1}^{2L} \sum_{m=-l+\frac{1}{2}}^{l-\frac{1}{2}}
\sum_{m'=-l'+\frac{1}{2}}^{l'-\frac{1}{2}} \frac{1}{l(l+1)}
\nonumber \\
&&\times \left[
{\text Tr}(Y_{l,m+\frac{1}{2}}^\dag \lambda 
Y_{l',m'+\frac{1}{2}})
{\text Tr}\left(
[L_+,Y_{l',m'+\frac{1}{2}}^\dag ] a_3 
[L_-,Y_{l,m+\frac{1}{2}}]  \right) \right. \nonumber \\
&&\left.
-{\text Tr}(Y_{l,m-\frac{1}{2}}^\dag \lambda 
Y_{l',m'-\frac{1}{2}})
{\text Tr}\left(
[L_-,Y_{l',m'-\frac{1}{2}}^\dag ] a_3 
[L_+,Y_{l,m-\frac{1}{2}}]  \right)
\right]. 
\label{I31}
\end{eqnarray}
Comparing with eq.(\ref{H31}), 
the only difference is the position of $\lambda$:
$\lambda$ is placed before the $Y_{lm}$ in (\ref{I31}), 
while it was after  $Y_{lm}$ in (\ref{H31}).
Then we can follow the same steps in $H_3$,
and obtain
\begin{eqnarray}
I&=&
\frac{-2ig^2\rho}{\alpha}
\sum_{n=0}^{\infty}\frac{1}{\{L(L+1)\}^{n+1}} \n
&&\times\Bigl[
{\text Tr}\left(L_{i_n}\cdots L_{i_1} \right)
\bigl[\epsilon_{ijk}{\text Tr}
\left(\lambda [L_i,a_j]L_kL_{i_1}\cdots L_{i_n} \right) 
-i{\text Tr}\left(\lambda a_i L_i  L_{i_1}\cdots L_{i_n} \right)\bigr] \n
&&-{\text Tr}\left(L_{i_n}\cdots L_{i_1} L_i \right)
\bigl[\epsilon_{ijk}{\text Tr}\left(\lambda [L_j,a_k]L_{i_1}\cdots L_{i_n} \right) 
-i{\text Tr}\left(\lambda a_i L_{i_1}\cdots L_{i_n} \right)\bigr]
\Bigr]. \n
\label{covanoterm}
\end{eqnarray}
Due to the position of $\lambda$ in (\ref{I31}), $\lambda$ and $a_i$ are
placed in the same trace, namely in the same integral when
we express traces by integrals. 
Therefore the WT identity can be written locally for an arbitrary
chiral parameter $\lambda$.
The WT identity (\ref{covWI1}) for the covariant current
becomes
\begin{equation}
\langle {\text Tr}
\left(\lambda\left[L_i+\rho a_i,-\psi_\beta L_j\bar\psi_\alpha 
(\sigma_i \sigma_j)^{\alpha\beta} \right] \right)
-\frac{2\rho^2}{\alpha}
{\text Tr}\left(\bar\psi \lambda \phi\psi \right) 
\rangle_S 
= I.
\label{covWI2}
\end{equation}
We will further evaluate the anomaly in two limiting cases below.
\subsubsection{Commutative Limit}
We now take the commutative limit.
In this limit, we can replace $L_i$ by the classical coordinate
$x_i/ \alpha$, and
${\cal L}_i$ by $\tilde{\cal L}_i$.
The second line of eq.(\ref{covanoterm}) becomes
\begin{eqnarray}
&&\frac{-2ig^2 \rho}{\alpha}\sum_{n=0}^\infty 
\frac{1}{\left\{L(L+1)\right\}^{2n+1}}
\frac{(2L+1)^2}{\alpha^{4n+1}}
\int_\Omega \left(
\epsilon_{ijk}\lambda [L_i,a_j ] x_k x_l^{2n}
-i\lambda (a\cdot x) x_l^{2n}\right)
\int_{\Omega'} (x'_l)^{2n} \nonumber \\
&=&
\frac{-2ig^2 \rho}{\alpha}\sum_{n=0}^\infty 
\frac{1}{\left\{L(L+1)\right\}^{2n+1}}
\frac{(2L+1)^2}{\alpha^{4n+1}}
\frac{\rho^{4n}}{2n+1}
\int_\Omega \left(
\epsilon_{ijk}\lambda [L_i,a_j ] x_k -i\lambda\rho\phi 
\right) \nonumber \\
&=&
\frac{-2ig^2}{\rho}(2L+1)^2
\sum_{n=0}^\infty\frac{1}{2n+1} \int_{\Omega}
\left(
\epsilon_{ijk}\lambda [L_i,a_j ] x_k -i\lambda\rho\phi 
\right).
\end{eqnarray}
Similarly the third line becomes
\begin{eqnarray}
&&\frac{2ig^2 \rho}{\alpha}\sum_{n=0}^\infty 
\frac{1}{\left\{L(L+1)\right\}^{2n+2}}
\frac{(2L+1)^2}{\alpha^{4n+3}}
\int_\Omega \left(
\epsilon_{ijk}\lambda [L_j,a_k ] x_l^{2n+1}
-i\lambda a_i x_l^{2n+1}\right)
\int_{\Omega'} (x'_l)^{2n+1} x'_i \nonumber \\
&=&
\frac{2ig^2 \rho}{\alpha}\sum_{n=0}^\infty 
\frac{1}{\left\{L(L+1)\right\}^{2n+2}}
\frac{(2L+1)^2}{\alpha^{4n+3}}
\frac{\rho^{4n+2}}{2n+3}
\int_\Omega \left(
\epsilon_{ijk}\lambda [L_i,a_j ] x_k -i\lambda \rho \phi
\right) \nonumber \\
&=&
\frac{2ig^2}{\rho}(2L+1)^2
\sum_{n=0}^\infty 
\frac{1}{(2n+3)}
\int_{\Omega}
\left(
\epsilon_{ijk}\lambda [L_i,a_j ] x_k -i\lambda \rho \phi
\right). 
\end{eqnarray}
Here we have used the following formula:
\begin{equation}
\int_\Omega x_{i_1} \cdots x_{i_{2n}} =
\frac{(\rho^2)^n}{(2n+1)!!}
\left[\delta_{i_1 i_2}\delta_{i_3 i_4}\cdots 
\delta_{i_{2n-1} i_{2n}}
+ \cdots  \right],
\end{equation}
where there are $(2n-1)!!$ ways of contraction in the 
r.h.s..
We thus obtain the commutative limit of $I$:
\begin{equation}
I\longrightarrow \frac{-2ig^2}{\rho}
(2L+1)^2
\int_\Omega \left(
\epsilon_{ijk}\lambda [L_i,a_j ] x_k -i\lambda\rho\phi 
\right).
\label{CL}
\end{equation}
The commutative limit of 
the WT identity (\ref{covWI2}) becomes
\begin{equation}
{\cal L}_i(\bar\psi \sigma_i \gamma_3 \psi)
-2\rho\phi\bar\psi\psi
-4g^2\phi
=\frac{4g^2}{\rho}(-i\epsilon_{ijk}x_i{\cal L}_j a_k-2\rho\phi),
\end{equation}
which completely agrees with  (\ref{comAWI1}).
The last term in the l.h.s.  is the normal-ordering
constant for the second term, as we have already seen 
in the noncommutative case at $\lambda=1$.
\subsubsection{Flat Limit}
We then take another limit of (\ref{covanoterm}), the flat limit.
Since we consider the vicinity of the north pole,
the background gauge field $a_i$ is assumed to have supports
only  around the north pole and
we can replace $L_i$ by $L_3$ if they are in the trace in which
$a_i$ and $\lambda$ are inserted:
\begin{eqnarray}
I &\rightarrow& \frac{-2ig^2\rho}{\alpha}\sum_{n=0}^\infty
\frac{1}{[L(L+1)]^{n+1}} \n
&&\times \Bigl[
{\text Tr}((L_3)^n)
[\epsilon_{ij}{\text Tr}(\lambda[L_i,a_j](L_3)^{n+1})
-i{\text Tr}(\lambda a_3(L_3)^{n+1})] \n
&&-{\text Tr}((L_3)^{n+1})
[\epsilon_{ij}{\text Tr}(\lambda [L_i,a_j](L_3)^n)
-i{\text Tr}(\lambda a_3 (L_3)^n)]
\Bigr].
\end{eqnarray}
Here $i,j$ take values 1 or 2.
By the same reasoning, $L_3$ in the trace in which $a_i$ is inserted
can be replaced with a c-number $\sqrt{L(L+1)}$.
${\text Tr}(L_3^n)$ can be evaluated exactly,
and it vanishes for odd $n$.
We then obtain
\begin{equation}
I \rightarrow
\frac{-4ig^2\rho}{\alpha}
\bigl[\epsilon_{ij}{\text Tr}(\lambda[L_i,a_j])
-i{\text Tr}(\lambda a_3)\bigr] ,
\end{equation}
and the WT identity (\ref{covWI2}) becomes
\begin{eqnarray}
&&\langle {\text Tr}
\left(\lambda\left[L_i+\rho a_i,-\psi L_3\bar\psi 
\sigma_i \sigma_3 \right] \right)
-\frac{2\rho^2}{\alpha}
{\text Tr}\left(\bar\psi \lambda \phi\psi \right) 
\rangle_S \n
&=&
\frac{-4ig^2\rho}{\alpha}
\bigl[\epsilon_{ij}{\text Tr}(\lambda[L_i,a_j])
-i{\text Tr}(\lambda a_3)\bigr].
\label{FL}
\end{eqnarray}
By using (\ref{flatint}), (\ref{flatadjLi}), 
(\ref{sigmapp}), (\ref{app}), we obtain
\begin{eqnarray}
&&\langle \int d^2 x
\left[ \lambda (i\partial_i +\tilde{a}''_i)
(-\sigma''_{i \alpha\beta}(\gamma''_{3 \beta\gamma} 
\psi^\gamma)\bar\psi^\alpha)
+\lambda[a_3,-\psi \bar\psi]
-2\bar\psi \lambda \phi\psi \right]_\star
\rangle_S \n
&=&
\int d^2 x
\bigl[  4g^2\lambda\epsilon_{ij}\partial_i a''_j
- \frac{1}{\rho}4g^2 \lambda a_3 \bigr]_\star,
\label{flatlimitcovWI}
\end{eqnarray}
where 
\begin{equation}
\gamma''_3=\frac{\alpha}{\rho}L^R_3 \sigma_3,
\end{equation}
$\tilde{a}''_i$ are adjoint operators of $a''_i$,
and $[\cdots]_\star$ means that the products in the bracket are
replaced by the star products.
Note that the last term of the r.h.s. in 
(\ref{flatlimitcovWI}) is a sub-leading contribution 
in $1/\rho$ 
and can be ignored.
The second term in the l.h.s. of (\ref{flatlimitcovWI}) 
vanishes up to the 
first order in the gauge field $a_i$,
since 
$
\langle \lambda[a_3,-\psi \bar\psi]
\rangle_{S_0}=0 
$
due to (\ref{psiApsiBs0}).
Therefore the WT identity becomes
\begin{equation}
\langle \int d^2 x
\left[ \lambda (i\partial_i +\tilde{a}''_i)
(-\sigma''_{i \alpha\beta}(\gamma''_{3 \beta\gamma} 
\psi^\gamma)\bar\psi^\alpha)
-2\bar\psi \lambda \phi\psi \right]_\star
\rangle_S 
=
\int d^2 x
\bigl[  4g^2\lambda\epsilon_{ij}\partial_i a''_j
 \bigr]_\star
\label{flatlimitcovWI2}  
\end{equation}
in the flat limit.

\section{Conclusions and Discussions}
\setcounter{equation}{0}
In this paper, we have calculated chiral anomaly for 
fermions in the fundamental representation 
on the fuzzy 2-sphere.
This system can be formulated as a matrix model of
finite size and no regularization is necessary.
In spite of this, we can reproduce the anomalous
chiral WT identity.
Our final results for the WT identities are written in 
eq.(\ref{WIinvfinal}) and eq.(\ref{covWI2}).
We have obtained WT identities for two types of chiral currents,
a gauge invariant current (\ref{WIinvfinal}) 
and a covariant current (\ref{covWI2}).

The anomaly term is contained in $H$ of eq.(\ref{invanoterm}) 
and $I$ of eq.(\ref{covanoterm}) respectively.
$H$ and $I$ have the same expression except the location 
of the chiral transformation parameter $\lambda$.
In $H$, $\lambda$ and the background gauge field $a_i$ 
are inserted in different traces, while in $I$, they 
are in the same trace.
${\text Tr}$ becomes an integral over 2-sphere. 
Hence, for the covariant case, if 
we take ${\text Tr}\lambda$ out of the WT 
identity (\ref{covWI2}) and the corresponding anomaly term 
$I$ in eq.(\ref{covanoterm}), we obtain 
a local expression for the anomaly term in the WT identity. 
On the contrary, for the invariant case, even after 
taking ${\text Tr}\lambda$ out of $H$, the gauge field 
$a_i$ is still in the other trace, namely in an integral, and 
the WT identity has a nonlocal form.
If we put $\lambda=1$, $H$ and $I$ become the same 
and we can obtain the same global form of the chiral anomaly. 

When we take a flat limit, the small difference between 
$H$ and $I$ causes a big difference to the final results.
By assuming that both of the chiral transformation 
parameter $\lambda$ and the gauge field $a_i$ are 
localized around the north pole of the sphere, the 
anomaly for the covariant case becomes a star generalization 
of the anomaly in the commutative theory as we saw 
in section 4.2.3. The anomaly for the invariant case, 
however, vanishes as in section 4.1.3. 
These results are consistent with the previous 
results\cite{AS1}-\cite{Martin:2001ye}. 
A merit of our calculation is that our system is finite 
and we could obtain the results (\ref{invanoterm}) 
and (\ref{covanoterm}), which interpolate 
between local $\lambda$ and global 
$\lambda$.

In this paper, we have only evaluated the anomaly in 
the leading order of the gauge field.
Though the calculation of higher orders is very complicated,
they can be guessed by the gauge covariance.
For the gauge invariant and covariant currents
respectively, 
$H$ of eq.(\ref{invanoterm}) 
and $I$ of eq.(\ref{covanoterm})
can
have the following simple gauge invariant completions: 
\begin{eqnarray}
&&H_G =  
-\frac{2i g^2 \rho^2}{\alpha}
\sum_{n=0}^{\infty} 
\frac{1}{\left\{ \alpha L(L+1)\right\}^{n+1}} \n 
&& \qquad \times \left[ {\text Tr}(L_{i_n}\cdots L_{i_1}\lambda )
{\text Tr}(A_i B_i A_{i_1}\cdots A_{i_n})  
-\alpha {\text Tr}(L_{i_n}\cdots L_{i_1}L_i \lambda )
{\text Tr}(B_i A_{i_1}\cdots A_{i_n}) 
\right] , \n 
&& \\
&&I_G = 
-\frac{2i g^2 \rho^2}{\alpha}
\sum_{n=0}^{\infty} 
\frac{1}{\left\{\alpha L(L+1)\right\}^{n+1}} \n 
&& \qquad \times \left[ {\text Tr}(L_{i_n}\cdots L_{i_1})
{\text Tr}(\lambda A_i B_i A_{i_1}\cdots A_{i_n}) 
-\alpha {\text Tr}(L_{i_n}\cdots L_{i_1}L_i)
{\text Tr}(\lambda B_i A_{i_1}\cdots A_{i_n}) 
\right] , \n
\end{eqnarray}
where  $B_i$ is
the magnetic field defined by 
\begin{eqnarray}
B_i&=&\frac{1}{2} \epsilon_{ijk}F_{jk} 
=\frac{1}{\alpha^2 \rho^2}
(\epsilon_{ijk}A_j A_k - i\alpha A_i), \\
F_{ij}&=&\frac{1}{\alpha^2 \rho^2}
\left([A_i, A_j]-i\alpha \epsilon_{ijk}A_k \right).
\end{eqnarray}
They are gauge covariant fields and vanish when
$a_i=0$. 
The above forms of anomalies are invariant under gauge 
transformations and become $H$ of (\ref{invanoterm}) 
and $I$ of (\ref{covanoterm})
in the first order of the gauge field.
In the commutative limit,
the anomaly for the covariant current becomes
the same as the r.h.s. of (\ref{CL}) except that
$[L_i, a_j]$ is replaced by a gauge covariant 
form $([L_i ,a_j]-[L_j, a_i]+\rho [a_i, a_j])/2$.
It is also similar in the flat limit, the final expression
becomes the same as the r.h.s. of eq.(\ref{FL})
except $[L_i, a_j]$ is replaced by the same covariant
form. 
For the case of the invariant current, it is interesting
to see whether the above gauge invariant completion 
is consistent with the perturbative
form of the anomaly discussed 
in \cite{Banerjee:2001un}. 
\par
A motivation to consider chiral anomaly on fuzzy 
2-sphere is to define topological invariants
on finite noncommutative space.
In commutative space, we have fully understood the 
topological structures of the gauge configuration
space and its relation to the index of Dirac operators
or anomalies.
They have been extensively utilized in many situations,
for example, in  constructing the chiral fermions
in the Kalza-Klein compactification. 
\par
We have been struggling to  build a constructive formulation
of superstrings and, at present, matrix models or
noncommutative field theories are considered to be promising
candidates. {}From this perspective, it is necessary
to investigate various possibilities to make chiral fermions
in finite dimensional matrix models. 
One possibility was investigated  in ref.\cite{OMM} where orbifold
matrix models were proposed.
Another interesting possibility to make chiral fermions
will be to define index of some
Dirac operators in a compactified noncommutative space
and then make use of the Dirac operator with a nontrivial index.
In infinite dimensional noncommutative space, 
solitons have been constructed\cite{NCS} 
in terms of the so-called shift operators.
But the shift operator is formally written as
\begin{equation}
 S=\sum_{n=0}^{\infty} |n+1 \rangle \langle n| ,
\end{equation}
and the construction  essentially makes use of the infinite
dimensionality. 
In finite noncommutative geometries, topologically nontrivial 
field configurations have been constructed based on algebraic 
K-theory and projective modules\cite{23_5,13_2}.
Though they are mathematically beautiful, it seems difficult 
to apply this idea to matrix models for square matrices, 
such as \cite{IKKT}.
\par
To overcome the above difficulty, 
it is interesting to apply the ideas related to 
Ginsparg--Wilson(GW) relation\cite{GinspargWilson} 
in lattice gauge theory
to matrix models or 
noncommutative field theories.
GW relation
represents the remnant chiral symmetry of chiral continuum 
theories.
Another important idea originates from the 
observation that 
in the presence of a mass defect, a chiral fermion appears
at the boundary. 
So far a domain wall fermion\cite{DW} and a
vortex fermion\cite{KN} are constructed on the lattice
and from the former model, a practical solution of GW relation 
is derived\cite{Neuberger}.
More abstractly, 
GW relation plays a crucial role in discussing 
the chiral symmetry\cite{Luscher,Nieder} or
the index theorem at a finite lattice 
spacing\cite{Hasenfratzindex,Luscher}.
In a forthcoming paper\cite{AIN} we show that
by making use of the ideas related to GW relation, it is
possible to define topological invariants
or indices of Dirac operators in the finite dimensional
fuzzy 2-sphere with general background gauge field 
configurations.

\section{Appendix}
\setcounter{equation}{0}
\subsection{Useful  Identities}
$L_i$'s are $(2L+1)$-dimensional representation
matrices of the  
angular momentum operators and satisfy
\begin{eqnarray}
[L_i,L_j]&=&i\epsilon_{ijk}L_k, \\
(L_i)^2&=&L(L+1), \\
L_\pm &=&L_1\pm iL_2, \\
\left[L_+,L_- \right]&=&2L_3, \\
\left[L_3 ,L_\pm \right]&=&\pm L_\pm.
\end{eqnarray}
${\cal L}_i M=[L_i,M]$ are adjoint operators and satisfy
\begin{equation}
[{\cal L}_i,{\cal L}_j]=i\epsilon_{ijk}{\cal L}_k.
\end{equation}
Noncommutative spherical harmonics $Y_{lm}$ satisfy
an orthonormality and a completeness relation:
\begin{eqnarray}
\frac{1}{2L+1}{\text Tr}(Y_{lm}^\dag Y_{l'm'})&=&
\delta_{ll'}\delta_{mm'}, \\
\frac{1}{2L+1}\sum_{l=0}^{2L} \sum_{m=-l}^l (Y_{lm}^\dag)_{ij}(Y_{lm})_{kp}&=&
\delta_{ip}\delta_{jk} \label{compY}.
\end{eqnarray}
Total number of basis wave functions becomes $\sum_{l=0}^{2L}(2l+1)=(2L+1)^2$ and
agrees with the number of independent $(2L+1)$-dimensional hermitian matrices.
Thus for any matrices $A$ and $B$, we have
\begin{eqnarray}
\frac{1}{2L+1}\sum_{lm}{\text Tr}(Y_{lm}^\dag A)
{\text Tr}(Y_{lm}B)&=&{\text Tr}(AB)\label{compYtt}, \\
\frac{1}{2L+1}\sum_{lm}{\text Tr}
(Y_{lm}^\dag A Y_{lm}B)&=&{\text Tr}(A){\text Tr}(B)\label{compYt}. 
\end{eqnarray}
There are various useful identities when $L_i$'s act
on $Y_{lm}$:
\begin{eqnarray}
\left[L_i,\left[L_i ,Y_{lm} \right]\right]&=&l(l+1) Y_{lm},\\
\left[L_3,Y_{lm}\right]&=& m Y_{lm}, \label{L3Y} \\
L_i Y_{lm} L_i &=& \left[L(L+1)-\frac{1}{2}l(l+1) \right]Y_{lm},\\
L_i \left[L_i,Y_{lm}\right]&=&\frac{l(l+1)}{2}Y_{lm},\\
\left[L_i,Y_{lm}\right]L_i &=&-\frac{l(l+1)}{2}Y_{lm},
\end{eqnarray}
\begin{eqnarray}
\left[L_+,Y_{lm}\right]&=& \sqrt{l(l+1)-m(m+1)}Y_{l,m+1},
\label{L+Y}\\
\left[L_-,Y_{lm}\right]&=& \sqrt{l(l+1)-m(m-1)}Y_{l,m-1}.
\label{L-Y}
\end{eqnarray}
Using the above equations, we can prove the following 
identity:
\begin{eqnarray}
\frac{1}{l(l+1)+2 \epsilon} Y_{lm}
&=&\sum_{n=0}^\infty \frac{1}{2L(L+1)}
\left[1-\frac{l(l+1)+2 \epsilon}{2L(L+1)}\right]^n Y_{lm},\n
&=&\sum_{n=0}^\infty \frac{1}{2L(L+1)}
\left[\frac{L_i^L L_i^R-\epsilon}{L(L+1)}\right]^n Y_{lm}.
\label{idll1}
\end{eqnarray}
\subsection{Expectation Values in the Free Theory}
The free part of the action $S_0$ (\ref{FuzzyAction}) becomes
\begin{eqnarray}
S_0
&=&\frac{\alpha (2L+1)}{2g^2}
\left[\sum_{l=0}^{2L} b_{l+\frac{1}{2},m}^\dag
b_{l+\frac{1}{2},m}(l+1)
- \sum_{l=1}^{2L} {b'}^\dag_{l-\frac{1}{2},m}
b'_{l-\frac{1}{2},m}l\right]
\end{eqnarray}
by expanding the fields $\psi$ and $\bar\psi$
in terms of spinorial-spherical harmonics 
${\cal Y}_{l+\frac{1}{2},m}$ and ${\cal Y'}_{l-\frac{1}{2},m}$ 
introduced in eq.(\ref{spinY}) as
\begin{eqnarray}
\psi&=&\sum_{l=0}^{2L}\sum_{m=-l-\frac{1}{2}}^{l+\frac{1}{2}}
b_{l+\frac{1}{2},m}{\cal Y}_{l+\frac{1}{2},m}
+
\sum_{l=1}^{2L}\sum_{m=-l+\frac{1}{2}}^{l-\frac{1}{2}}
b'_{l-\frac{1}{2},m}{\cal Y'}_{l-\frac{1}{2},m}, \\
\bar\psi &=&\sum_{l=0}^{2L}
\sum_{m=-l-\frac{1}{2}}^{l+\frac{1}{2}}
b_{l+\frac{1}{2},m}^\dag {\cal Y}^\dag_{l+\frac{1}{2},m}
+
\sum_{l=1}^{2L}\sum_{m=-l+\frac{1}{2}}^{l-\frac{1}{2}}
{b'}^\dag_{l-\frac{1}{2},m}{\cal Y'}^\dag_{l-\frac{1}{2},m}.
\end{eqnarray}
Then we can calculate the following expectation values,
\begin{eqnarray}
\langle O \rangle_{S_0}&=&\frac{1}{Z_{S_0}}
\int d\psi d\bar\psi O e^{-S_0} ,\\
Z_{S_0}&=&\int d\psi d\bar\psi e^{-S_0} ,
\end{eqnarray}
by using the Wick's theorem. 
Thus, we obtain the
following formulae:
\begin{eqnarray}
&\bullet& \langle {\text Tr}(\bar\psi A\psi B) \rangle_{S_0}
=-\frac{4g^2}{\alpha(2L+1)}{\text Tr}(AB) .
\label{psiApsiBs0}\\
&\bullet& \langle {\text Tr}(\bar\psi A \sigma_3 \psi B) \rangle_{S_0}=
-\frac{4g^2}{\alpha(2L+1)}
\sum_{l=1}^{2L}\sum_{m=-l}^{l}
\frac{m}{l(l+1)} {\text Tr}(Y_{lm}^\dag A Y_{lm} B) . \\
&\bullet& \langle {\text Tr}(\bar\psi A \sigma_{\pm} \psi B) \rangle_{S_0}=
-\frac{2g^2}{\alpha(2L+1)}
\sum_{l=1}^{2L}\sum_{m=-l+\frac{1}{2}}^{l-\frac{1}{2}}
\frac{\sqrt{\left(l+\frac{1}{2}\right)^2 -m^2}}{l(l+1)} 
{\text Tr}\left(Y_{l,m\mp\frac{1}{2}}^\dag A Y_{l,m\pm\frac{1}{2}} B\right) . \nonumber \\
\end{eqnarray}
\begin{eqnarray}
&\bullet & \langle {\text Tr}(\bar\psi A\psi B)
{\text Tr}(\bar\psi C\psi D) \rangle_{S_0} \nonumber \\
&&=
\frac{8g^4}{\alpha^2 (2L+1)^2}
{\text Tr}(AB){\text Tr}(CD)\n
&&-\frac{8g^4}{\alpha^2 (2L+1)^2}
\sum_{l=1}^{2L}\sum_{l'=1}^{2L}\sum_{m=-l}^{l}
\sum_{m'=-l'}^{l'} \frac{mm'}{l(l+1)l'(l'+1)}
{\text Tr}(Y_{lm}^\dag A Y_{l'm'} B){\text Tr}
(Y_{l'm'}^\dag C Y_{lm} D)\n
&&-\frac{4g^4}{\alpha^2(2L+1)^2}
\sum_{l=1}^{2L}\sum_{l'=1}^{2L}
\sum_{m=-l+\frac{1}{2}}^{l-\frac{1}{2}}
\sum_{m'=-l'+\frac{1}{2}}^{l'-\frac{1}{2}} \n 
&&\times \frac{1}{l(l+1)l'(l'+1)}
\sqrt{\left(l+\frac{1}{2}\right)^2-m^2}
\sqrt{\left(l'+\frac{1}{2}\right)^2-{m'}^2} \nonumber \\
&& \times\left[ 
{\text Tr}(Y_{l,m-\frac{1}{2}}^\dag A 
Y_{l',m'-\frac{1}{2}} B){\text Tr}
(Y_{l',m'+\frac{1}{2}}^\dag C Y_{l,m+\frac{1}{2}} D)\right.\n
&&\left.+{\text Tr}(Y_{l,m+\frac{1}{2}}^\dag A 
Y_{l',m'+\frac{1}{2}} B){\text Tr}
(Y_{l',m'-\frac{1}{2}}^\dag C Y_{l,m-\frac{1}{2}} D)
\right]. 
\end{eqnarray}
\begin{eqnarray}
&\bullet& \langle {\text Tr}(\bar\psi A\psi B)
{\text Tr}(\bar\psi C \sigma_3 \psi D) \rangle_{S_0}
\nonumber \\
&&=
\frac{16g^4}{\alpha^2 (2L+1)^2}
\sum_{l=1}^{2L}\sum_{m=-l}^{l} \frac{m}{l(l+1)}
{\text Tr}(AB){\text Tr}(Y_{lm}^\dag C Y_{lm} D)\n
&&-\frac{4g^4}{\alpha^2 (2L+1)^2}
\sum_{l=1}^{2L}\sum_{l'=1}^{2L}
\sum_{m=-l+\frac{1}{2}}^{l-\frac{1}{2}}
\sum_{m'=-l'+\frac{1}{2}}^{l'-\frac{1}{2}} \n
&& \times \frac{1}{l(l+1)l'(l'+1)}
\sqrt{\left(l+\frac{1}{2}\right)^2-m^2}
\sqrt{\left(l'+\frac{1}{2}\right)^2-{m'}^2} \nonumber \\
&&\times
\left[
{\text Tr}(Y_{l,m+\frac{1}{2}}^\dag A 
Y_{l',m'+\frac{1}{2}} B){\text Tr}
(Y_{l',m'-\frac{1}{2}}^\dag C Y_{l,m-\frac{1}{2}} D)\right.\n
&&
\left.-
{\text Tr}(Y_{l,m-\frac{1}{2}}^\dag A 
Y_{l',m'-\frac{1}{2}} B){\text Tr}
(Y_{l',m'+\frac{1}{2}}^\dag C Y_{l,m+\frac{1}{2}} D)
\right] \n
&&-
\frac{8g^4}{\alpha^2(2L+1)^2}
\sum_{l=1}^{2L}\sum_{m=-l}^{l}
\frac{m}{l(l+1)}
\left[{\text Tr}(Y_{lm}^\dag A B){\text Tr}(C Y_{lm} D)
+{\text Tr}(A Y_{lm} B){\text Tr}(Y_{lm}^\dag C D)
\right]. \nonumber \\
\label{psiApsiBpsiCsigma3psiDs0}
\end{eqnarray}
\begin{eqnarray}
&\bullet& \langle {\text Tr}(\bar\psi A\psi B)
{\text Tr}(\bar\psi C \sigma_\pm \psi D) \rangle_{S_0}
\nonumber \\
&&=
\frac{8g^4}{\alpha^2 (2L+1)^2}
\sum_{l=1}^{2L}\sum_{m=-l+\frac{1}{2}}^{l-\frac{1}{2}} 
\frac{\sqrt{\left(l+\frac{1}{2}\right)^2 -m^2}}{l(l+1)}
{\text Tr}(AB)
{\text Tr}(Y_{l,m\mp\frac{1}{2}}^\dag C Y_{l,m\pm\frac{1}{2}} D)\n
&& \pm \frac{4g^4}{\alpha^2 (2L+1)^2}
\sum_{l=1}^{2L}\sum_{l'=1}^{2L} \nonumber \\
&& \times
\left[           
\sum_{m=-l}^{l}
\sum_{m'=-l'+\frac{1}{2}}^{l'-\frac{1}{2}}
\frac{m\sqrt{\left(l'+\frac{1}{2}\right)^2-{m'}^2}}{l(l+1)l'(l'+1)}
{\text Tr}(Y_{lm}^\dag A 
Y_{l',m'\pm \frac{1}{2}} B){\text Tr}
(Y_{l',m'\mp \frac{1}{2}}^\dag C Y_{lm} D)\right.\nonumber \\
&&\left.
-
\sum_{m=-l+\frac{1}{2}}^{l-\frac{1}{2}}
\sum_{m'=-l'}^{l'}
\frac{m'\sqrt{\left(l+\frac{1}{2}\right)^2-m^2}}
{l(l+1)l'(l'+1)}
{\text Tr}(Y_{l,m\mp \frac{1}{2}}^\dag A 
Y_{l'm'} B){\text Tr}
(Y_{l'm'}^\dag C Y_{l,m\pm \frac{1}{2}} D) \right] \nonumber \\
&&-
\frac{4g^4}{\alpha^2(2L+1)^2} \times
\left[
\sum_{l=1}^{2L}\sum_{m=-l+\frac{1}{2}}^{l-\frac{1}{2}}
\frac{\sqrt{\left(l+\frac{1}{2}\right)^2-m^2}}{l(l+1)}
{\text Tr}(Y_{l,m\mp\frac{1}{2}}^\dag A B)
{\text Tr}(C Y_{l,m\pm\frac{1}{2}} D) \right.\nonumber\\
&& \left. +
\sum_{l'=1}^{2L}\sum_{m'=-l'+\frac{1}{2}}^{l'-\frac{1}{2}}
\frac{\sqrt{\left(l'+\frac{1}{2}\right)^2-{m'}^2}}{l'(l'+1)}
{\text Tr}(A Y_{l',m'\pm \frac{1}{2}} B)
{\text Tr}(Y_{l',m'\mp\frac{1}{2}}^\dag C D)
\right]. 
\end{eqnarray}
Here $A,B,C,D$ are arbitrary $(2L+1)$-dimensional matrices and 
$\sigma_{\pm}=\frac{\sigma_1 \pm i\sigma_2}{2}$.

\subsection{Chiral Anomaly on $S^2$}\label{sec:comWI}
In this appendix, we will derive 
anomalous WT identity in commutative 2-sphere,
(\ref{comAWI1}),({\ref{comAWI2}).
Under the chiral transformation (\ref{comCT}),
the action (\ref{comAction}) varies as 
\begin{eqnarray}
\delta S_{S^2} = \frac{\rho}{g^2}\int_{\Omega} \left[ 
(\tilde{\cal L}_i \lambda) \bar\psi \sigma_i \gamma_3 \psi 
+2\rho\lambda \bar\psi\phi\psi \right],
\end{eqnarray}
and the measure changes $d\psi'd\bar\psi'=Jd\psi\bar\psi$.
The Jacobian is calculated as 
\begin{eqnarray}
J&=&\exp\left[-2\int_\Omega \lambda(x) 
\sum_n \phi_n^\dag (x) \gamma_3 \phi_n (x) \right] \\
&=&1-2 \int_\Omega \lambda(x) 
\sum_n \phi_n^\dag (x) \gamma_3 \phi_n (x), 
\end{eqnarray}
where $\phi_n$ is a complete set of eigenfunctions for 
the hermitian Dirac operator $D$ and 
satisfies $D\phi_n =\lambda_n\phi_n$ and 
$\int_{\Omega}\phi_n^\dag(x)\phi_n(x)=\delta_{nm}$.

Here we introduce 
%
spherical harmonics $Y_{lm}$, which satisfy
\begin{eqnarray}
\int_{\Omega} Y_{lm}^* (x) Y_{l'm'}(x)&=&
\delta_{ll'}\delta_{mm'}, \\
\sum_{l=0}^\infty \sum_{m=-l}^{m=l}
Y_{lm} (x) Y_{lm}^*(x')&=& 4\pi \delta(\Omega-\Omega'),\\
\sum_{m=-l}^{m=l}
Y_{lm}^*(x') Y_{lm} (x) &=&(2l+1)P_l(\cos\alpha), \\
\sum_{m=-l}^{m=l}
Y_{lm}^*(x) Y_{lm} (x) &=&2l+1, \label{YY}\\
\sum_{m=-l}^{m=l}
Y_{lm}^*(x) \tilde{\cal L}_iY_{lm} (x) &=&0, \label{YLY}\\
\sum_{m=-l}^{m=l}
Y_{lm}^*(x)\tilde{\cal L}_i\tilde{\cal L}_j Y_{lm} (x) 
&=&\frac{(2l+1)l(l+1)}{2}[\delta_{ij}-\frac{x_ix_j}{\rho^2}],
\label{YLLY}
\end{eqnarray}
where $\alpha$ in the third line is the angle between 
$\Omega$ and $\Omega'$.

We then evaluate the Jacobian using Fujikawa's method\cite{Fujikawa} as
\begin{eqnarray}
{\cal A}(x)
&=&\sum_n \phi_n^\dag (x) \gamma_3 \phi_n (x) \n
&=&\lim_{M\rightarrow\infty}\sum_n 
\phi_n^\dag (x) \gamma_3 \exp\left(-\frac{\lambda_n^2}{M^2}\right)\phi_n (x) \n
&=&\lim_{M\rightarrow\infty}
\sum_{l=0}^{\infty} \sum_{m=-l}^{l}\sum_{s =\pm 1} 
\chi_s^\dagger Y_{lm}^* (x) \gamma_3 \exp\left(-\frac{D^2}{M^2}\right)
Y_{lm} (x) \chi_s\n
&=&\lim_{M\rightarrow\infty}
\sum_{l=0}^{\infty} \sum_{m=-l}^{l} e^{-\frac{l(l+1)}{M^2}}
Y_{lm}^* (x) {\text tr} \left[\gamma_3 \exp\left(-\frac{D'}{M^2}\right)\right]
Y_{lm} (x), \label{YDprimeY}
\end{eqnarray}
where $\chi_s$ in the third line
are complete basis for the spinor space.
In the last line, ${\text tr}$ is a trace over spinor space, and
\begin{eqnarray}
D'&=&D^2-(\tilde{\cal L}_i)^2 \n
&=&\sigma_i\tilde{\cal L}_i+1
+\rho[(\tilde{\cal L}_i a_i)
+i\epsilon_{ijk}\sigma_k(\tilde{\cal L}_i a_j) 
+2\sigma_i a_i  
+2a_i \tilde{\cal L}_i ]
+\rho^2a_ia_i.\label{Dprime}
\end{eqnarray}
Only the first term $\sigma_i\tilde{\cal L}_i$ and the sixth 
term $2\rho a_i \tilde{\cal L}_i$ in eq.(\ref{Dprime}) can act 
as angular momentum operators 
on the factors in the right in eq.(\ref{YDprimeY}),
and can take the value of the order of $l$, and thus $M$,
when acting on $Y_{lm}$.
The other $\tilde{\cal L}_i$'s in eq.(\ref{Dprime})
act only on $a_i$ in the round bracket.
Thus, the terms except for the first and the sixth terms
in eq.(\ref{Dprime})
act just as c-number,
and take the value of the order of $1$ in eq.
(\ref{YDprimeY}).
Therefore, when we Taylor-expand 
$\exp\left(-\frac{D'}{M^2}\right)$, 
only the following terms  
can survive in the large $M$ limit in eq.(\ref{YDprimeY}): 
\begin{equation}
{\text tr}\left[\gamma_3\exp\left(-\frac{D'}{M^2}\right) \right]
\to 
{\text tr}\left[\gamma_3\left(1-\frac{D'}{M^2}
+\frac{1}{2M^4}
(\sigma_i\tilde{\cal L}_i+2\rho a_j \tilde{\cal L}_j)^2
\right)\right].
\end{equation}
After taking trace over the spinor space, this becomes
\begin{equation}
\frac{-2}{M^2}[x_i \tilde{\cal L}_i/\rho 
+i\epsilon_{ijk} x_k(\tilde{\cal L}_i a_j)+2x_i a_i]
+\frac{4}{M^4}x_ia_j\tilde{\cal L}_i\tilde{\cal L}_j.
\end{equation}
By using (\ref{YY}),(\ref{YLY}),(\ref{YLLY}),
\begin{equation}
{\cal A}(x)=
\lim_{M\rightarrow\infty}
\sum_{l=0}^{\infty} e^{-\frac{l(l+1)}{M^2}}(2l+1)
(\frac{-2}{M^2})[i\epsilon_{ijk} x_k(\tilde{\cal L}_i a_j)+2x_i a_i].
\end{equation}
The summation over the variable $l$ can be 
transfered to the integral of the continuous variable $l$ as, 

\begin{eqnarray}
{\cal A}(x)&=&M\int_0^{\infty} dl e^{-l^2} (2Ml) 
(\frac{-2}{M^2})[i\epsilon_{ijk} x_k(\tilde{\cal L}_i a_j)+2x_i a_i]\\
&=&2(-i\epsilon_{ijk} x_k (\tilde{\cal L}_i a_j) -2\phi \rho )\\
&=&2\rho\epsilon_{ijk}x_i\partial_j a'_k,
\end{eqnarray}
where $a'_i $
and $\phi$ 
are defined in eqs.(\ref{aprime}), (\ref{scalarphi}). 

Using the above results, the WT identity, 
$\langle \delta S_{S^2} +2\int_{\Omega}\lambda {\cal A} \rangle=0 $, 
is written as 
\begin{eqnarray}
 &&\frac{\rho}{g^2}\langle \int_{\Omega} \lambda(x)(
 \tilde{\cal L}_i (\bar\psi \sigma_i \gamma_3 \psi) 
-2\rho\bar\psi 
\phi\psi) \rangle_S \n
&=& \int_{\Omega} \lambda(x) \left[ 
-4i\epsilon_{ijk} x_k (\tilde{\cal L}_i a_j) 
-8\phi\rho \right]  \\
&=& \int_{\Omega} \lambda(x) \left[ 
4 \rho\epsilon_{ijk} x_i \partial_j a'_k \right], 
\end{eqnarray}
which gives (\ref{comAWI1}) or (\ref{comAWI2}).

\subsection{Evaluation for $\lambda=1$ Case}
\label{H3lambda1}
In this appendix, we evaluate $H_3 |_{\lambda=1}$ in (\ref{H3lambda12}).
By making use of the identities 
(\ref{L3Y}) and (\ref{idll1}), we have
\begin{eqnarray}
H_3 |_{\lambda=1} 
&=&\frac{4g^2\rho}{\alpha(2L+1)}\sum_{l=0}^{2L}\sum_{m=-l}^l
\sum_{n=0}^\infty \frac{1}{2L(L+1)} 
\frac{1}{\left\{L(L+1) \right\}^n} \n
&&\times{\text Tr}\left(Y_{lm}^\dag a_3 (L_i^L L_i^R)^n 
\left[L_3,Y_{lm} \right] \right) \label{epsilonzero} \nonumber\\
&=&
\frac{4g^2\rho}{\alpha}\frac{1}{2L(L+1)}
\sum_{n=0}^\infty  \frac{1}{\left\{L(L+1) \right\}^n}\nonumber\\
&&\times 
\left[{\text Tr}\left( a_3 L_{i_n}\cdots L_{i_1} L_3 \right) 
{\text Tr}\left( L_{i_1}\cdots L_{i_n} \right) 
-{\text Tr}\left( a_3 L_{i_n}\cdots L_{i_1} \right) 
{\text Tr}\left( L_3 L_{i_1}\cdots L_{i_n} \right) 
\right] \nonumber \\
&\equiv&
\frac{4g^2\rho}{\alpha}\frac{1}{2L(L+1)}
\sum_{n=0}^\infty f_n. \label{Ca1}
\end{eqnarray}
In the first line, $l=0$ term is added in the sum since 
$[L_3, Y_{00}]=0$. In the end of this appendix, we will justify
it more carefully.
Several $f_n$ with small $n$ can be evaluated as
\begin{eqnarray}
f_0&=&(2L+1){\text Tr}(a_3 L_3), \\
f_1&=&-\frac{1}{3} (2L+1) {\text Tr}(a_3 L_3), \\
f_2&=&\frac{1}{3} (2L+1) {\text Tr}(a_3 L_3) 
-\frac{1}{6L(L+1)} (2L+1) {\text Tr}(a_3 L_3), \\
f_3&=&\frac{1}{6L(L+1)} (2L+1) {\text Tr}(a_3 L_3)\nonumber \\
&&-\frac{2L+1}{15 \left\{ L(L+1)\right\}^2 }
\left[ L(L+1)(2L^2+2L+1)+(L^2+L-1)(L^2+L-2) \right]\n
&&\times{\text Tr}(a_3 L_3). 
\end{eqnarray}
Thus, the sum $\sum f_n$ is proportional to ${\text Tr} (a_3 L_3)$:
\begin{equation}
\sum_{n=0}^\infty f_n=C{\text Tr(a_3 L_3)}. \label{Ca2}
\end{equation}
In order to evaluate the value of $C$, we replace 
the index $3$ by a general index $i$ and sum over $i$.
We then set  $a_i=L_i$. 
Then the r.h.s. of eq.(\ref{Ca2}) becomes 
\begin{equation}
C{\text Tr(a_3 L_3)}\vert_{3\rightarrow i,a_i\rightarrow L_i} 
= CL(L+1)(2L+1). \label{Ca3}
\end{equation}
On the other hand, from eq.(\ref{Ca1}) we obtain 
\begin{eqnarray}
&&\sum_{n=0}^\infty f_n \vert_{3\rightarrow i,a_i\rightarrow L_i}\n
&=&
\sum_{n=0}^\infty \frac{1}{\left\{ L(L+1)\right\}^n} 
\Bigl[
L(L+1)
{\text Tr}\left(L_{i_n}\cdots L_{i_1} \right) 
{\text Tr}\left( L_{i_1}\cdots L_{i_n} \right) \n
&&\hspace{3cm}-{\text Tr}\left(L_{i_n+1} L_{i_n}\cdots L_{i_1}\right) 
{\text Tr}\left( L_{i_1}\cdots L_{i_n}L_{i_n+1} \right) 
\Bigr]\nonumber  \\
&=&
L(L+1)\left\{{\text Tr}1 \right\}^2
-\lim_{n\rightarrow\infty}\frac{1}{\left\{ L(L+1)\right\}^n}
\left\{{\text Tr}(L^{n+1}) \right\}^2 \nonumber\\
&=&
4\left\{ L(L+1)\right\}^2, \label{Ca4}
\end{eqnarray}
where we have used the following identity
\begin{eqnarray}
&&\lim_{n\rightarrow\infty}\frac{1}{\left\{ L(L+1)\right\}^n}
{\text Tr}\left(L_{i_n+1}\cdots L_{i_1} \right) 
{\text Tr}\left( L_{i_1}\cdots L_{i_n+1} \right) \n
&=&
\lim_{n\rightarrow\infty}\frac{1}{\left\{ L(L+1)\right\}^n}
\frac{1}{2L+1}\sum_{lm}
{\text Tr}[Y_{lm}^\dagger L_{i_n+1}\cdots L_{i_1}  
Y_{lm} L_{i_1}\cdots L_{i_n+1}]\n
&=&
\lim_{n\rightarrow\infty}\frac{1}{\left\{ L(L+1)\right\}^{n+1}}
\sum_{lm}
[L(L+1)-\frac{1}{2}l(l+1)]^n \n
&=&L(L+1). 
\end{eqnarray}
{}From eqs.(\ref{Ca2}),(\ref{Ca3}),(\ref{Ca4}), we find  
\begin{equation}
\sum_{n=0}^\infty f_n=\frac{4L(L+1)}{2L+1} {\text Tr}(a_3 L_3).
\end{equation}
Thus, from eq.(\ref{Ca1}), we obtain
\begin{equation}
H_3 |_{\lambda=1}=\frac{8g^2\rho}{\alpha(2L+1)} 
{\text Tr}(a_3 L_3).
\end{equation}

In the remainder of this appendix, we 
treat $l=0$ part more carefully by introducing a
regulator $\epsilon$
and justify the above calculation.
Eq.(\ref{epsilonzero}) becomes
\begin{eqnarray}
&&\frac{4g^2\rho}{\alpha(2L+1)}\sum_{l=0}^{2L}\sum_{m=-l}^l
\sum_{n=0}^\infty \frac{1}{2L(L+1)} 
\frac{1}{\left\{L(L+1) \right\}^n} \n
&&\times{\text Tr}\left(Y_{lm}^\dag a_3 
\left[L_3,(L_i^L L_i^R-\epsilon)^n 
Y_{lm} \right] \right).
\end{eqnarray}
Then, after replacing $3$ by $i$ ($i$ is summed) and setting $a_i=L_i$,
we have
\begin{eqnarray}
&&\sum_{n=0}^\infty f_n \vert_{3\rightarrow i,a_i\rightarrow L_i}\n
&=&\frac{1}{(2L+1)}\sum_{lmn}
\frac{1}{\left\{L(L+1) \right\}^n} 
\biggl[L(L+1)
{\text Tr}\left(Y_{lm}^\dag 
(L_i^L L_i^R-\epsilon)^n Y_{lm}\right)\n
&&\hspace{5cm}-{\text Tr}\left(Y_{lm}^\dag 
L_i^L L_i^R(L_i^L L_i^R-\epsilon)^n Y_{lm}\right)
\biggr]\n
&=&\frac{1}{(2L+1)}\biggl[
L(L+1)(2L+1)^3 \n
&&\hspace{2cm}
-\lim_{n\rightarrow\infty}
\frac{1}{\left\{L(L+1) \right\}^n}
\sum_{lm} {\text Tr}\left(Y_{lm}^\dag 
(L_i^L L_i^R-\epsilon)^{n+1} Y_{lm}\right) \n
&&\hspace{2cm}
-\sum_{lmn}\frac{\epsilon}{\left\{L(L+1) \right\}^n} 
{\text Tr}\left(Y_{lm}^\dag 
(L_i^L L_i^R-\epsilon)^n Y_{lm}\right)
\biggr]\n
&=&\frac{1}{(2L+1)}\biggl[
L(L+1)(2L+1)^3
-0
-2L(L+1)(2L+1)\biggr]\n
&=&4[L(L+1)]^2,
\end{eqnarray}
which exactly agrees with (\ref{Ca4}).


\begin{thebibliography}{99}
\bibitem{SW} N.~Seiberg and E.~Witten,
JHEP {\bf 9909}, 032 (1999)
[arXiv:hep-th/9908142].
\bibitem{CDS} A.~Connes, M.~R.~Douglas and A.~Schwarz,
JHEP {\bf 9802}, 003 (1998)
[arXiv:hep-th/9711162].
\bibitem{NCMM}
H.~Aoki, N.~Ishibashi, S.~Iso, H.~Kawai, Y.~Kitazawa and T.~Tada,
Nucl.\ Phys.\ B {\bf 565}, 176 (2000)
[arXiv:hep-th/9908141].
\bibitem{MLi}
M.~Li,
Nucl.\ Phys.\ B {\bf 499}, 149 (1997)
[arXiv:hep-th/9612222].
%

\bibitem{NC}D.~Bigatti and L.~Susskind,
Phys.\ Rev.\ D {\bf 62}, 066004 (2000)
[arXiv:hep-th/9908056];

N.~Ishibashi, S.~Iso, H.~Kawai and Y.~Kitazawa,
Nucl.\ Phys.\ B {\bf 573}, 573 (2000)
[arXiv:hep-th/9910004];  

S.~Minwalla, M.~Van Raamsdonk and N.~Seiberg,
JHEP {\bf 0002}, 020 (2000)
[arXiv:hep-th/9912072];
\par
S.~Iso, H.~Kawai and Y.~Kitazawa,
Nucl.\ Phys.\ B {\bf 576}, 375 (2000)
[arXiv:hep-th/0001027]; 
\par
A.~Matusis, L.~Susskind and N.~Toumbas,
JHEP {\bf 0012}, 002 (2000)
[arXiv:hep-th/0002075];
\par
N.~Ishibashi, S.~Iso, H.~Kawai and Y.~Kitazawa,
Nucl.\ Phys.\ B {\bf 583}, 159 (2000)
[arXiv:hep-th/0004038];
\par
N.~Seiberg,
JHEP {\bf 0009}, 003 (2000)
[arXiv:hep-th/0008013].
\par
\bibitem{IKKT}
N.~Ishibashi, H.~Kawai, Y.~Kitazawa and A.~Tsuchiya,
Nucl.\ Phys.\ B {\bf 498}, 467 (1997)
[arXiv:hep-th/9612115].
\par
For a review, see
H.~Aoki, S.~Iso, H.~Kawai, Y.~Kitazawa, A.~Tsuchiya and T.~Tada,
Prog.\ Theor.\ Phys.\ Suppl.\  {\bf 134}, 47 (1999)
[arXiv:hep-th/9908038].

\bibitem{AS1} F.~Ardalan and N.~Sadooghi,
Int.\ J.\ Mod.\ Phys.\ A {\bf 16}, 3151 (2001)
[arXiv:hep-th/0002143].
\bibitem{Martin} J.~M.~Gracia-Bondia and C.~P.~Martin,
Phys.\ Lett.\ B {\bf 479}, 321 (2000)
[arXiv:hep-th/0002171].
\bibitem{AS2}F.~Ardalan and N.~Sadooghi,
Int.\ J.\ Mod.\ Phys.\ A {\bf 17}, 123 (2002)
[arXiv:hep-th/0009233].

\bibitem{Banerjee:2001un}
It was pointed out, however, that gauge invariant form of 
anomaly can be constructed in a perturbative expansion 
with respect to the noncommutative parameter $\theta$.
See, 
R.~Banerjee and S.~Ghosh,
Phys.\ Lett.\ B {\bf 533}, 162 (2002)
[arXiv:hep-th/0110177].

\bibitem{Armoni} A.~Armoni, E.~Lopez and S.~Theisen,
JHEP {\bf 0206}, 050 (2002)
[arXiv:hep-th/0203165].
\bibitem{Nakajima} T.~Nakajima,
arXiv:hep-th/0205058.



\bibitem{Bonora:2000he}
L.~Bonora, M.~Schnabl and A.~Tomasiello,
Phys.\ Lett.\ B {\bf 485}, 311 (2000)
[arXiv:hep-th/0002210].

\bibitem{Moreno:2000xu}
E.~F.~Moreno and F.~A.~Schaposnik,
JHEP {\bf 0003}, 032 (2000)
[arXiv:hep-th/0002236].


\bibitem{Moreno:2000kt}
E.~F.~Moreno and F.~A.~Schaposnik,
Nucl.\ Phys.\ B {\bf 596}, 439 (2001)
[arXiv:hep-th/0008118].


\bibitem{Martin:2000qf}
C.~P.~Martin,
J.\ Phys.\ A {\bf 34}, 9037 (2001)
[arXiv:hep-th/0008126].


\bibitem{Grisaru:2000sk}
M.~T.~Grisaru and S.~Penati,
Phys.\ Lett.\ B {\bf 504}, 89 (2001)
[arXiv:hep-th/0010177].


\bibitem{Nishimura:2001dq}
J.~Nishimura and M.~A.~Vazquez-Mozo,
JHEP {\bf 0108}, 033 (2001)
[arXiv:hep-th/0107110].


\bibitem{Bonora:2001fa}
L.~Bonora and A.~Sorin,
Phys.\ Lett.\ B {\bf 521}, 421 (2001)
[arXiv:hep-th/0109204].


\bibitem{Martin:2001ye}
C.~P.~Martin,
Nucl.\ Phys.\ B {\bf 623}, 150 (2002)
[arXiv:hep-th/0110046].


\bibitem{13_1}
H.~Grosse and P.~Presnajder,
arXiv:hep-th/9805085;
%
Lett.\ Math.\ Phys.\  {\bf 46}, 61 (1998);

P.~Presnajder,
J.\ Math.\ Phys.\  {\bf 41}, 2789 (2000)
[arXiv:hep-th/9912050];

\bibitem{13_2}
A.~P.~Balachandran and S.~Vaidya,
Int.\ J.\ Mod.\ Phys.\ A {\bf 16}, 17 (2001)
[arXiv:hep-th/9910129];

\bibitem{13_3}
G.~Immirzi and B.~Ydri,
arXiv:hep-th/0203121.

\bibitem{KS} L.~H.~Karsten and J.~Smit,
Nucl.\ Phys.\ B {\bf 183}, 103 (1981).

\bibitem{nielsen}
H.~B.~Nielsen and M.~Ninomiya,
Nucl.\ Phys.\ B {\bf 185}, 20 (1981)
[Erratum-ibid.\ B {\bf 195}, 541 (1982)];
{\it ibid}\ B {\bf 193}, 173 (1981);
Phys.\ Lett.\ B {\bf 105}, 219 (1981).

\bibitem{IKTW}
S.~Iso, Y.~Kimura, K.~Tanaka and K.~Wakatsuki,
Nucl.\ Phys.\ B {\bf 604}, 121 (2001)
[arXiv:hep-th/0101102].


\bibitem{Grosse}H.~Grosse and J.~Madore,
Phys.\ Lett.\ B {\bf 283}, 218 (1992);

H.~Grosse and P.~Presnajder,
Lett.\ Math.\ Phys.\  {\bf 33}, 171 (1995);

H.~Grosse, C.~Klimcik and P.~Presnajder,
Commun.\ Math.\ Phys.\  {\bf 185}, 155 (1997)
[arXiv:hep-th/9507074];
%
arXiv:hep-th/9603071.
   
\bibitem{Watamura} 
U.~Carow-Watamura and S.~Watamura,
Commun.\ Math.\ Phys.\  {\bf 183}, 365 (1997)
[arXiv:hep-th/9605003].

\bibitem{bala} A.~P.~Balachandran, T.~R.~Govindarajan and B.~Ydri,
Mod.\ Phys.\ Lett.\ A {\bf 15}, 1279 (2000)
[arXiv:hep-th/9911087];
arXiv:hep-th/0006216.
%

\bibitem{stein}
C.~S.~Chu, J.~Madore and H.~Steinacker,
JHEP {\bf 0108}, 038 (2001)
[arXiv:hep-th/0106205];

\bibitem{wong}
P.~J.~O'Donnell and B.~Wong,
Phys.\ Lett.\ B {\bf 138}, 274 (1984).

\bibitem{OMM}H.~Aoki, S.~Iso and T.~Suyama,
Nucl. Phys. B634 (2002) 71
[arXiv:hep-th/0203277].

\bibitem{NCS} For a review, see J.~Harvey,
arXiv:hep-th/0102076.
%

\bibitem{23_5}
H.~Grosse, C.~Klimcik and P.~Presnajder,
Commun.\ Math.\ Phys.\  {\bf 178}, 507 (1996)
[arXiv:hep-th/9510083];

S.~Baez, A.~P.~Balachandran, B.~Idri and S.~Vaidya,
Commun.\ Math.\ Phys.\  {\bf 208}, 787 (2000)
[arXiv:hep-th/9811169];

G.~Landi,
J.\ Geom.\ Phys.\  {\bf 37}, 47 (2001)
[arXiv:math-ph/9905014].

%
\bibitem{GinspargWilson}P.~H.~Ginsparg and K.~G.~Wilson,
Phys.\ Rev.\ D {\bf 25}, 2649 (1982).

%
\bibitem{DW}D.~B.~Kaplan,
Phys.\ Lett.\ B {\bf 288}, 342 (1992)
[arXiv:hep-lat/9206013];

Y.~Shamir,
Nucl.\ Phys.\ B {\bf 406}, 90 (1993)
[arXiv:hep-lat/9303005];

V.~Furman and Y.~Shamir,
Nucl.\ Phys.\ B {\bf 439}, 54 (1995)
[arXiv:hep-lat/9405004].

%
\bibitem{KN}K.~Nagao,
Nucl.\ Phys.\ B {\bf 636}, 264 (2002)
[arXiv:hep-lat/0112030].

%
\bibitem{Neuberger}H.~Neuberger,
Phys.\ Lett.\ B {\bf 417}, 141 (1998)
[arXiv:hep-lat/9707022];

Phys.\ Rev.\ D {\bf 57}, 5417 (1998)
[arXiv:hep-lat/9710089];

Phys.\ Lett.\ B {\bf 427}, 353 (1998)
[arXiv:hep-lat/9801031].

 %
\bibitem{Luscher}M.~L\"uscher,
Phys.\ Lett.\ B {\bf 428}, 342 (1998)
[arXiv:hep-lat/9802011].

 %
\bibitem{Nieder}F.~Niedermayer,
Nucl.\ Phys.\ Proc.\ Suppl.\  {\bf 73}, 105 (1999)
[arXiv:hep-lat/9810026].

%
\bibitem{Hasenfratzindex}P.~Hasenfratz,
Nucl.\ Phys.\ Proc.\ Suppl.\  {\bf 63}, 53 (1998)
[arXiv:hep-lat/9709110];

P.~Hasenfratz, V.~Laliena and F.~Niedermayer,
Phys.\ Lett.\ B {\bf 427}, 125 (1998)
[arXiv:hep-lat/9801021].

\bibitem{AIN}H.~Aoki, S.~Iso and K.~Nagao, arXiv:hep-th/0209223.

 %
\bibitem{Fujikawa}K.~Fujikawa,
Phys.\ Rev.\ Lett.\  {\bf 42}, 1195 (1979);
Phys.\ Rev.\ D {\bf 21}, 2848 (1980)
[Erratum-ibid.\ D {\bf 22}, 1499 (1980)].









\end{thebibliography}
\end{document}